\documentclass{elsarticle}

\usepackage{lineno,hyperref}
\usepackage{float} 
\usepackage[caption=false]{subfig}
\usepackage{graphicx}
\usepackage{amsmath}
\usepackage{booktabs}
\usepackage{multirow}
\usepackage[numbers]{natbib}
\usepackage{caption}
\usepackage{url}
\usepackage{paralist}

\journal{no choice made yet}
\makeatletter
\def\ps@pprintTitle{%
 \let\@oddhead\@empty
 \let\@evenhead\@empty
 \def\@oddfoot{}%
 \let\@evenfoot\@oddfoot}
\makeatother

\begin{document}

\begin{frontmatter}

\title{Predicting Stochastic Travel Times based on High-Volume Floating Car Data}

\author[a]{Rodrigo Gon\c{c}alves}
\author[b]{Rui J. de Almeida}
\author[c]{Remco M. Dijkman}

\address[a]{ASML, The Netherlands}
\address[b]{Maastricht University, The Netherlands}
\address[c]{Eindhoven University of Technology, The Netherlands}

\begin{abstract}
Transportation planning depends on predictions of the travel times between loading and unloading locations. While accurate techniques exist for making deterministic predictions of travel times based on real-world data, making stochastic predictions remains an open issue. This paper aims to fill this gap by showing how floating car data from TomTom can be used to make stochastic predictions of travel times. It also shows how these predictions are affected by choices that can be made with respect to the level of aggregation of the data in space and time, and by choices regarding the dependence between travel times on different parts of the route. \end{abstract}

\end{frontmatter}

\section{Introduction}\label{sec:intro}

In an European context, improvement of route and travel information is a key strategy point in the Dutch Mobility Policy. According to \cite{dutchpolicy}, travellers should arrive at their destination on time in 95\% of the cases. During peak hours, travelling times must not be more than 150\% of the travelling time outside of peak times on motorways, and it must not be more than 200\% on urban ring roads and roads other than motorways that are managed by the government. Reliable traffic information is essential for the development of efficient traffic control, management and planning strategies. It includes the observation of three fundamental variables: flow, density and speed - also known as a traffic state \cite{daganzo1997fundamentals}.

However, in practice transportation companies often find that day-to-day travel times do not correspond to predicted travel times, which leads to frequent changes to the transportation plans that they made. At the same time the travel time predictions that are used, are deterministic, while the agreements that they have with their customers are usually stochastic in nature; customers accept, for example, a 95\% within delivery window performance, or impose a fine for late deliveries. Using reliable and stochastic travel times can help to improve resource utilization, reduce slack time, and reduce the number of changes to the transportation plan.

Therefore, the goal of this paper is to show the different options for aggregating stochastic travel times from real-world floating car data, based on the floating car data database of TomTom. A floating car data databases contain historical data on the time that it took individual cars to travel particular stretches of road, also called `links'. We focus on real-world floating car data, because of the high level of detail in which travel time information can be extracted from such data compared to other sources of travel times.

To predict stochastic travel times, we need to:
\begin{compactenum}
\item (technologically) extract the floating car data from the database;
\item aggregate probability distributions of travel times from the floating car data; and
\item determine the probability distribution that most accurately predicts the travel time for a particular route on a particular date and time in the future.
\end{compactenum}
With respect to the aggregation of probability distributions, we will show how choices concerning the level of aggregation affect the accuracy of the produced probability distribution. These choices concern, among others, the timeframe from which the historical data will be taken, whether historical data will be aggregated by route or by link, and the level of time-dependence of travel times that is used. Similarly, the choice for the probability distribution that best fits a the travel time of a planned route (say the route from Rotterdam to Venlo tomorrow at 9:00) affects the accuracy of the probability distribution. We can choose to use a distribution based on the same day last week, the entire last week, the same days in the past four weeks, etc. We will show the consequences of these choices.

Against this background, the remainder of this paper is structured as follows. Section~\ref{sec:sota} presents the state of the art on travel time prediction. Section~\ref{sec:data} presents the floating car data database from which we aggregate our distributions. Section~\ref{sec:speed} and section~\ref{sec:timedependence} present a descriptive analysis of the data, showing how floating car data can be aggregated into speed fields that visualize the time and space dependent speed that can be realized by cars traveling a particular link, and showing the extent to which travel speed - and consequently travel times - are time-dependent. Section~\ref{sec:estimate} discusses how travel time distributions can be estimated from floating car data, and Section~\ref{sec:prediction} discusses how distributions based on historical floating car data can be used to predict future travel times. Finally, section~\ref{sec:conclusions} presents the conclusions.

\section{State of the Art}\label{sec:sota}

There is a long history of research on techniques for predicting travel times and traffic speeds. Traditionally, these techniques are based on creating traffic-flow models that model the characteristics of a particular road network~\cite{Hoogendoorn2001state}. Subsequently, these models can be used to predict travel speeds and times on a particular part of the network. However, as an increasing amount of data is being collected on historical travel times, data-driven techniques are increasing in popularity. Hybrid techniques that combine model-driven and data-driven techniques exist as well~\cite{vanLint2005advanced}. Data-driven techniques for travel time prediction can be distinguished from each other based on a number of criteria.

First, the underlying analysis techniques that are used to make the travel time prediction differ greatly. Typically, it is possible to distinguish between parametric techniques, that are based on statistical analysis, and non-parametric techniques that are based on machine learning. Popular in the parametric category are time-series based techniques~\cite{Min2011real, Williams2001multivariate, Ishak2002performance} and Bayesian techniques~\cite{ Chen2014real, Fei2011bayesian}. In the non-parametric category, neural networks are popular~\cite{vanLint2005advanced,Dia2001object,Hinsbergen2011bayesian}, but other techniques are used as well~\cite{Elhenawy2014dynamic}.

Second, the timeframe for which the prediction is made differs. Most techniques focus on short-term (also called real-time) predictions~\cite{Nanthawichit2003application,vanLint2005advanced,Chen2014real,Fei2011bayesian,Min2011real}. For transportation planning, long-term predictions~\cite{Elhenawy2014dynamic} are more useful, because transportation plans are usually made one or more days before a trip is executed.

Third, the manner in which uncertainty in the travel time is addressed differs. Many papers make deterministic predictions. However, there also is a body of research that purely focuses on modeling the uncertainty itself~\cite{Kim2014finite,Susilawati2014urban, Kaparias2008new,Eisele2015estimating} and, of course, papers that incorporate uncertainty in the prediction itself~\cite{Yeon2008breakdown,Ramezani2012estimation,Rahmani2015floating,Caceres2016estimating}. Uncertainty is usually considered in the form of a lower and upper bound to the travel time at a certain level of confidence. To the best of our knowledge, no data-based distributions (in the form of histograms) have been considered.

Fourth, the assumed source of the historical travel times differs. Loop detectors are often used~\cite{vanLint2005advanced, Min2011real, Fei2011bayesian, Chen2014real, Sun2008travel}. The drawback of loop detectors is that they the entire network, such that the results of an analysis with loop detectors is only applicable to the parts of the network on which they are installed. Alternatively, floating car data may be used~\cite{Seo2015estimation,Nanthawichit2003application,Rahmani2015floating}. While floating car data can potentially provide information on the entire network, the data on each individual road may be sparse. Hybrid techniques that use both loop detectors and floating car data are also being developed~\cite{Qiu2010estimation,Ou2008piecewise,Elhenawy2014dynamic}. Meanwhile, companies like Google and TomTom are reaching an install base that is large enough to have historical data on a large part of the international road network. For many roads the dataset that is being created is even large enough to create usable time-dependent and stochastic predictions.

This work primarily relates to work on predicting travel times based on floating-car data. However, we use a particularly dense source of floating car data, while existing work uses less dense sources. For that reason, the primary challenge of existing work is to accurately predict travel times given that there is limited data available to do so. We solve a different challenge: given an extensive dataset of floating car data, how much data is necessary and sufficient to create a prediction that is as accurate as possible?

\section{Floating car data}\label{sec:data}

The growing need of the driving public for accurate traffic information has spurred the deployment of large scale dedicated monitoring infrastructure systems, which mainly consist in the use of inductive loop detectors and video cameras. \cite{herrera2010evaluation}. Such observation methods can acquire all traffic state variables but they are confined to the location where they were installed, making it unpractical to cover wide-ranging areas. Floating car data, on the other hand, provides a cost-effective way to collect wide-ranging traffic data by making use of the existent communication infrastructure. Floating car data is taken as moving sensors, travelling in a traffic flow while sending GPS data.
\begin{figure}[tbh]
	\centering
	\captionsetup{margin=1cm}
	\includegraphics[width=0.85\textwidth]{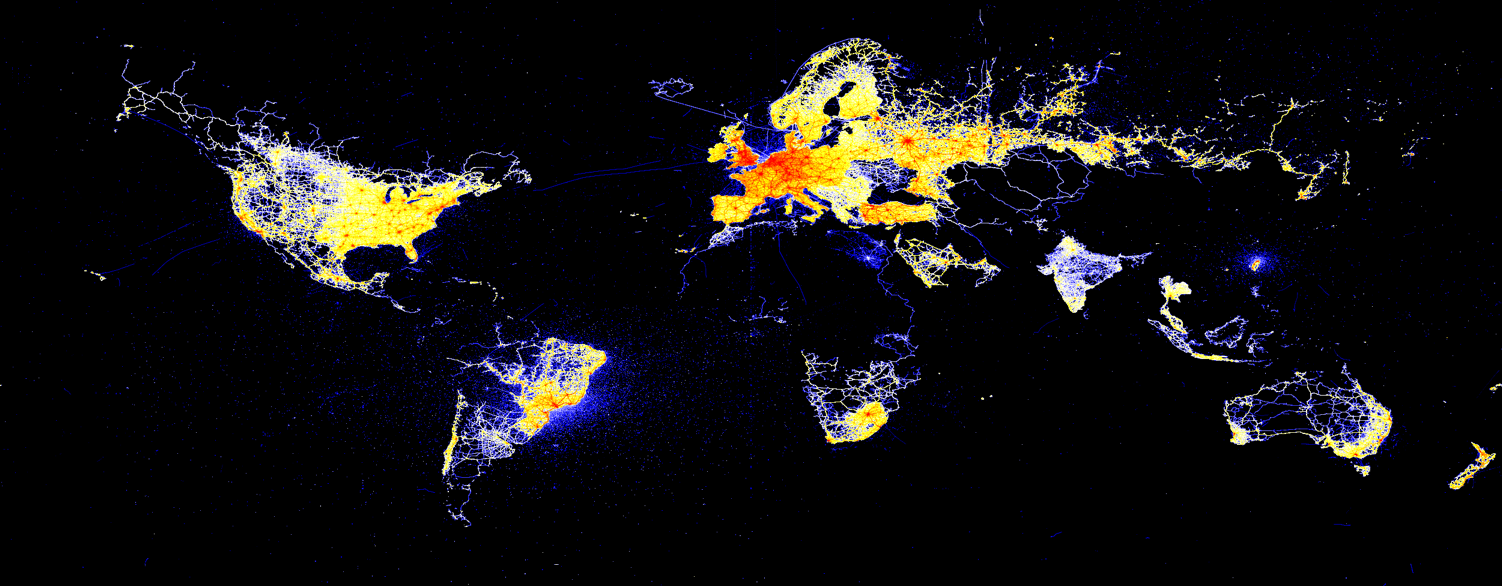}
	\caption{ TomTom probe data collected in Q3 of 2015}\label{tomtomcoverage}
\end{figure}
The TomTom Custom Travel Times API provides highly accurate speed and bottleneck information for individual road segments, which are called \emph{links}, using historical traffic data. Links are defined as the road segment between two road intersections. Because of that, link lengths can vary from a couple of meters, in urban areas, to several kilometers on highways. This is a key difference when compared to dual loop detector data where, normally, detectors are installed with a fixed distance between them. The TomTom data covers all roads, from major motorways to local and destination routes, throughout Europe and North America and a number of other countries including South Africa, Australia and New Zealand. The historical traffic information collected gives valuable insight into the traffic situation on the road network throughout the day. The API provides both route-based and link-based statistics for a route defined between two locations and, optionally, via waypoints. While link-based statistics concern the information gathered on road segment level, route-based statistics are related to the overall length of the trip, from origin to destination. However, clearly more data will be available per link than per route, as it is less likely that a car traveled the exact same route than the exact same link (i.e. part of a route).

A request for floating car data from the TomTom Custom Travel Times API can carry the following parameters.
\begin{compactitem}
\item \textbf{request type and location} - whether the distribution is requested for a route from a location $A$ to a location $B$, or for a link from a crossing $A$ to a crossing $B$;
\item \textbf{date range} - the period used for data collection;
\item \textbf{days of week} (DOW) - the days of the week used for data collection (e.g. only Mondays; all working days); 
\item \textbf{time of day} (TOD) - the period of the day used for data collection (e.g. from 10h00 to 12h00) and;
\item \textbf{full traversal} - which, if enabled, only allows vehicles that traversed the entire route to be used for data collection. Otherwise, all vehicles will contribute, even if they traversed only a small percentage of the route.
\end{compactitem}
While the ``full traversal'' option is useful to evaluate short routes or turn movements on junctions, it has a big impact on sample sizes, especially for uncommon trips, where sample sizes can drop to zero. For this reason, the ``full traversal'' parameter was kept disabled.

A request can return the following information:
\begin{compactitem}
\item Covered distance;
\item Sample size;
\item Travel time percentiles $\left(5^{th}, 10^{th}, \ldots, 90^{th}, 95^{th}\right)$;
\item Speed percentiles  $\left(5^{th}, 10^{th}, \ldots, 90^{th}, 95^{th}\right)$; and
\item Speed limit (only for link-based requests).
\end{compactitem}

\section{Travel speeds and times}\label{sec:speed}
A detailed representation of the traffic state in space and time allows us to analyse various aspects of traffic dynamics \cite{treiber2013traffic}. \citeauthor{treiber2013traffic} described a reconstruction of the full traffic state by spatio-temporal interpolation of the available traffic data, in a subset of locations and time. The \textit{adaptive smoothing method} (ASM) is a two-dimensional spatio-temporal interpolation algorithm to estimate the \textit{speed field} - a continuous function of local median speed  $V(x,t)$ given only discrete speed measurements $v_i$ at discrete locations $x_i$ and times $t_i$. Notice that, unlike in the original description of the method, we used local median speed rather than average to better cope with skewed distributions due to outliers. Empirical investigations of spatio-temporal traffic features has shown that traffic can be either ``free'' or ``congested'' \cite{kerner2012physics}. An obvious pattern observed in real traffic flows: the higher the vehicle density, the lower the average vehicle speed. \citeauthor{treiber2013traffic} distinguish these traffic states by observing how perturbations propagate in the traffic flow. In \textit{free traffic}, perturbations usually move down-stream at a characteristic velocity slightly below the local speed of the vehicles. In \textit{congested traffic}, perturbations propagate against the traffic flow (upstream) due to the reaction of the drivers to their respective leading vehicles. To account for these fundamental properties, two discrete convolutions, with a kernel $\phi_0$ and different propagation velocities in free and congested traffic, $c_{free}$ and $c_{cong}$, are considered:
\begin{equation}\label{Vfree}
{V_{free}(x,t)=\frac{1}{\mathcal{N}_{free}(x,t)}\sum_{i}^{ }\phi_0 \left(x-x_i,t-t_i-\frac{x-x_i}{c_{free}}\right)v_i}
\end{equation}
\begin{equation}\label{Vcong}
V_{cong}(x,t)=\frac{1}{\mathcal{N}_{cong}(x,t)}\sum_{i}^{ }\phi_0 \left(x-x_i,t-t_i-\frac{x-x_i}{c_{cong}}\right)v_i
\end{equation}
For the weighting kernel $\phi_0$, we used the same function as in \cite{treiber2013traffic}. However, unlike in dual loop detector data, distances between measurements are not equally spaced in floating car data. Due to this, the smoothing width in the spatial coordinates, $\sigma$, is defined has half of the link's $l_i$ length: $\sigma_i=\Delta l_i/2$. This condensates all speed information, obtained via GPS throughout the link, in a single point in space: the middle point of the link. The temporal smoothing width, $\tau$, is given by half of the TOD interval length. Parameters used in this experiment are summarized in Table \ref{asm parameters table}. The denominator $\mathcal{N}$ of Eqs. \ref{Vfree} and \ref{Vcong} denotes the normalization of the weighting function and is formulated as in \cite{treiber2013traffic}.
\begin{equation} 
\phi_0(x-x_i,t-t_i)=exp\left[-\left(\frac{\left|x-x_i\right|}{\sigma_i}+\frac{\left|t-t_i\right|}{\tau}\right)\right]
\end{equation}
The final speed field $V(x,t)$ is a combination of the two speed fields $V_{free}$ and $V_{cong}$, weighted by the ``degree of congestion'', $w$, also defined as in \cite{treiber2013traffic}.
\begin{equation}
 V(x,t)=\omega(x,t)V_{cong}(x,t)+\left[1-\omega(x,t)\right]V_{free}(x,t)
\end{equation}
\begin{figure}[H]
	\centering
	\captionsetup{margin=1cm}
	\subfloat[\label{discrete speed field}]{
		\includegraphics[width=0.45\textwidth]{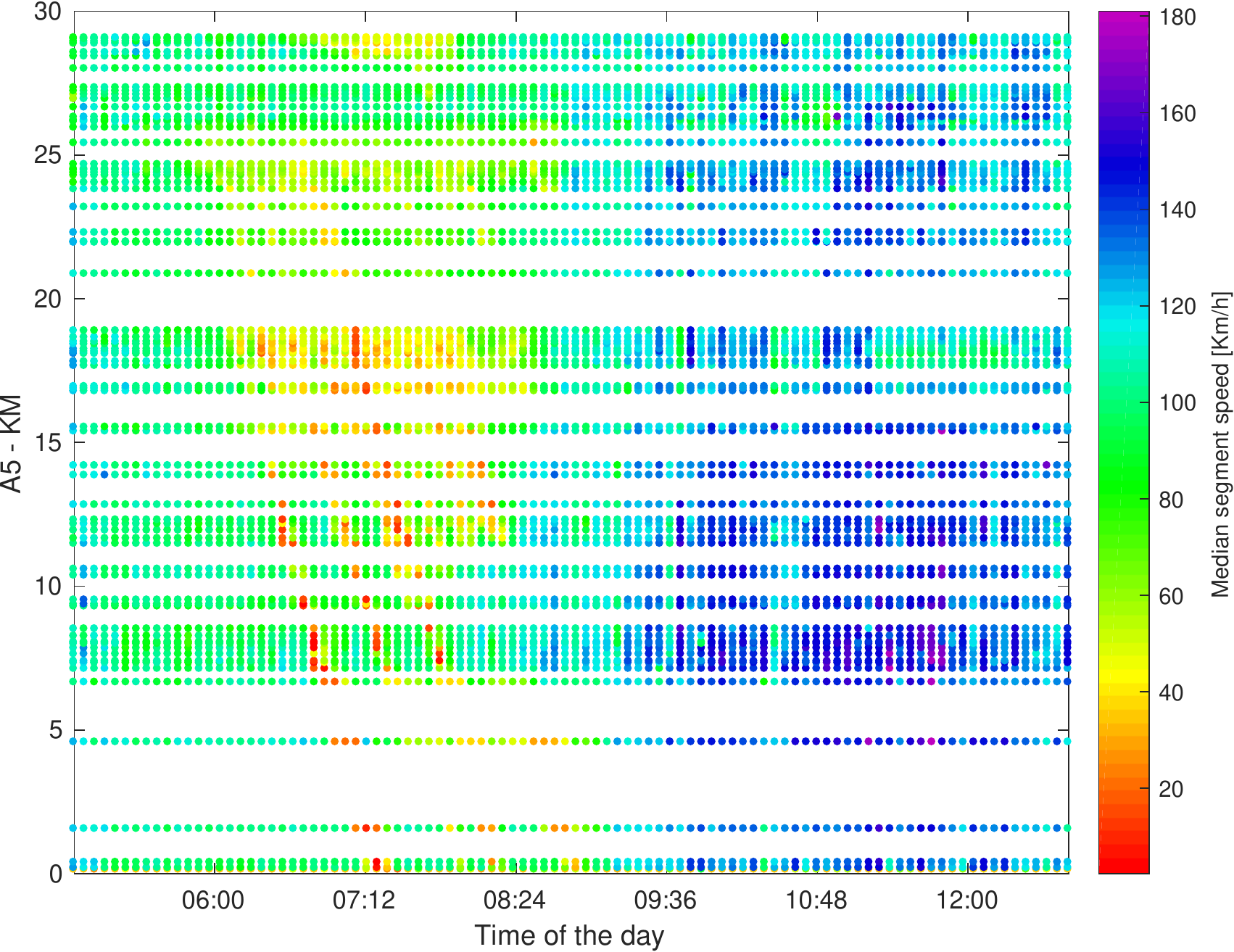}}
	\hspace{\fill}
	\subfloat[\label{continuous speed field}]{
		\includegraphics[width=0.45\textwidth]{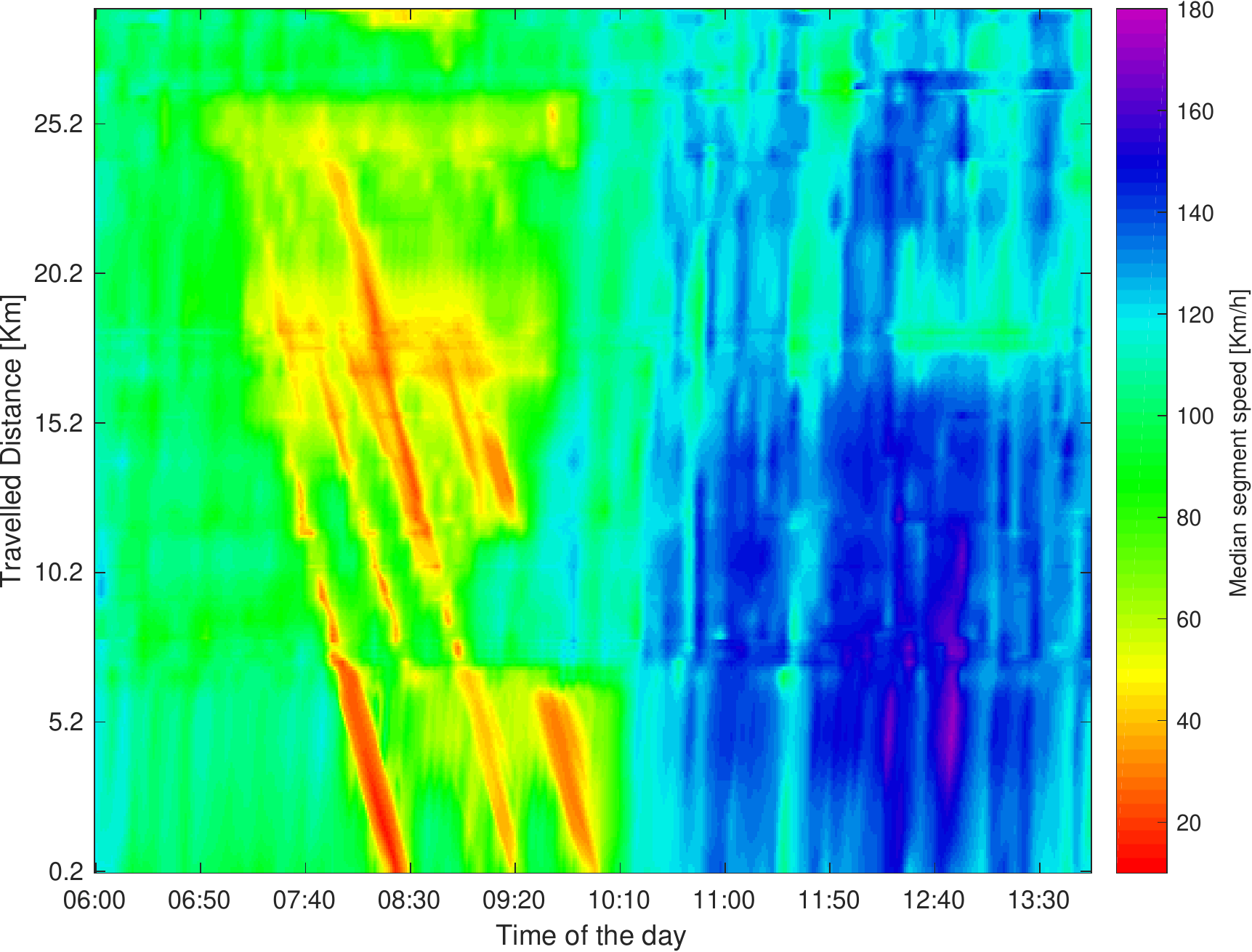}}
	\caption{ Colour-coded visualization of the mean speed obtained via TomTom API (a); and the same speed field reconstructed by the ASM (b).}\label{fig: Adaptive Smothing Method}
\end{figure}
As described in section \ref{sec:data}, data is available in the form of cumulative speed distributions, for each link of the specified route, aggregated over a TOD interval. Each point in fig.~\ref{discrete speed field}, represents the median speed $\left(50^{th}\text{ percentile}\right)$ on link $l_i$ collected on a 5 minute time interval. Uneven link lengths are the reason for the uneven white spaces in between measurements. Fig.~\ref{continuous speed field} is the resultant continuous speed field after interpolation with the \textit{adaptive smoothing method}. Traffic-state reconstruction methods using interpolation techniques are very useful for offline analysis of historical traffic data. Knowing the local speed field $V(x,t)$ it is relatively easy to generate virtual trajectories, by assuring the vehicle speed is identical to the corresponding local speed, and evaluate possible delays by changing departure time. These provide a good indicator of the total delay caused by congestion in which road managers and national economists are also interested.
\begin{figure}[H]
	\centering
	\captionsetup{margin=1cm}
	\subfloat[\label{monday1} \scriptsize{Monday, 1$^{st}$ February}]{
		\includegraphics[width=0.45\textwidth]{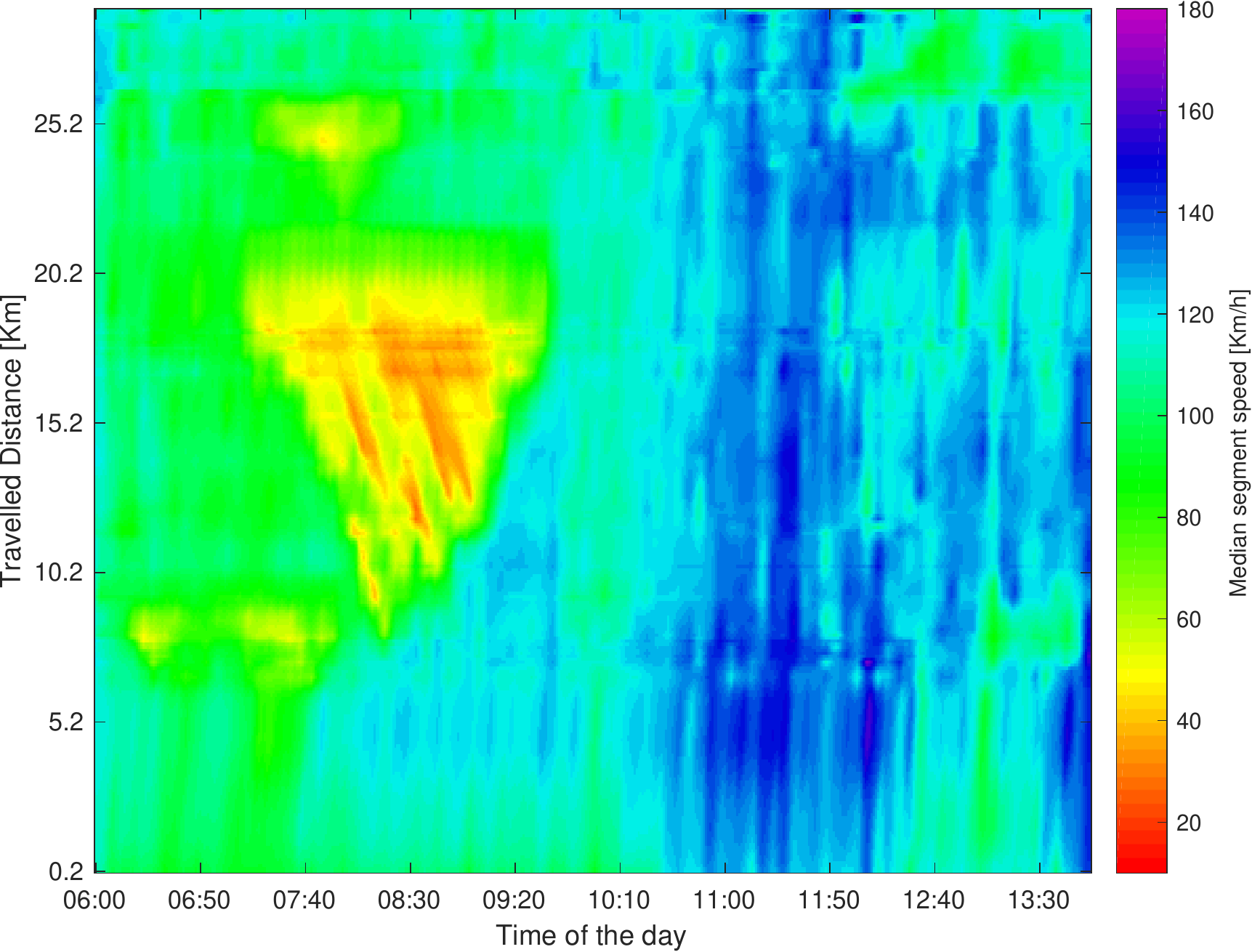}}
		\hspace*{\fill}
	\subfloat[\label{monday2} \scriptsize Monday, 8$^{th}$ February]{
		\includegraphics[width=0.45\textwidth]{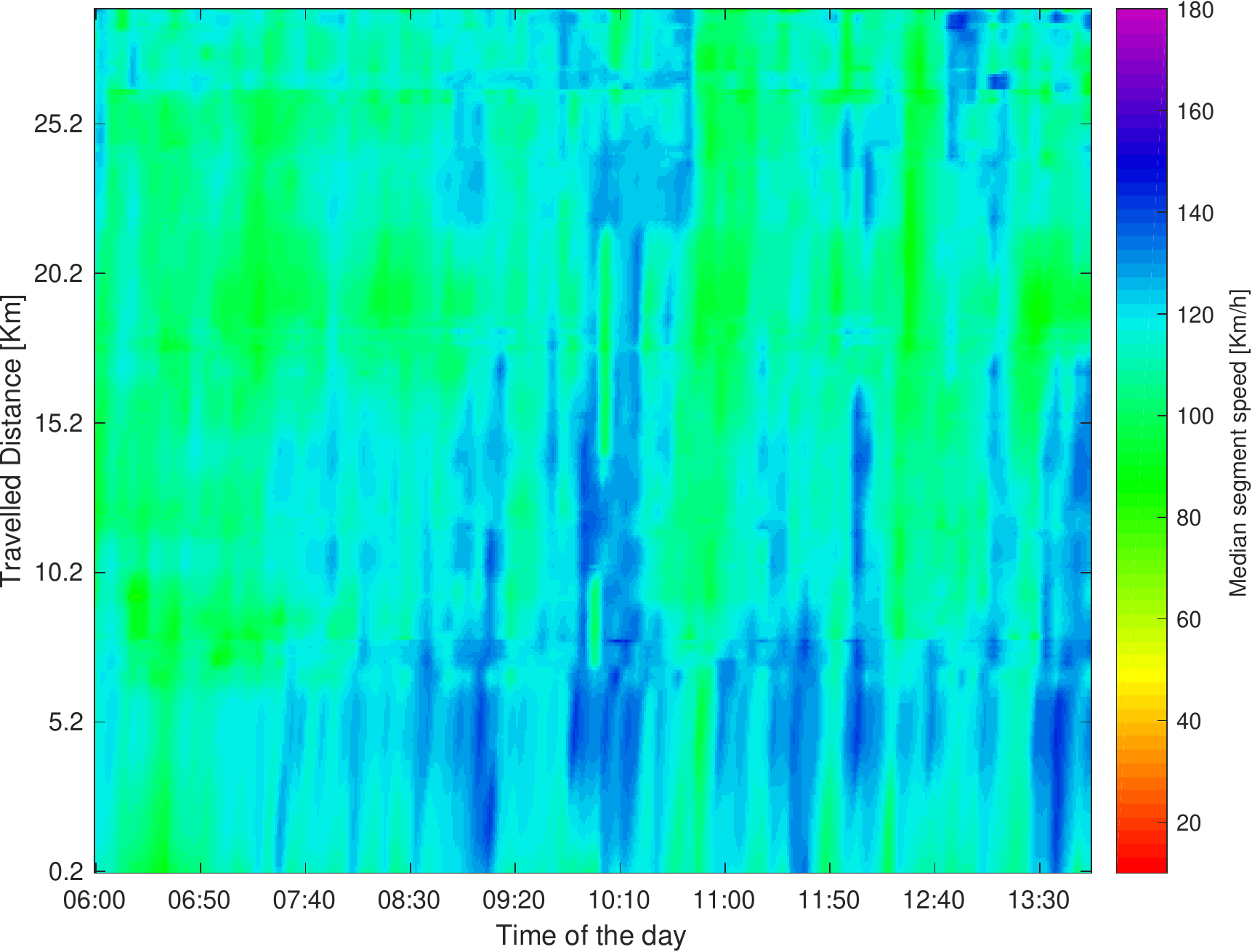}}\\
	\subfloat[\label{monday3} \scriptsize{Monday, 15$^{th}$ February}]{
		\includegraphics[width=0.45\textwidth]{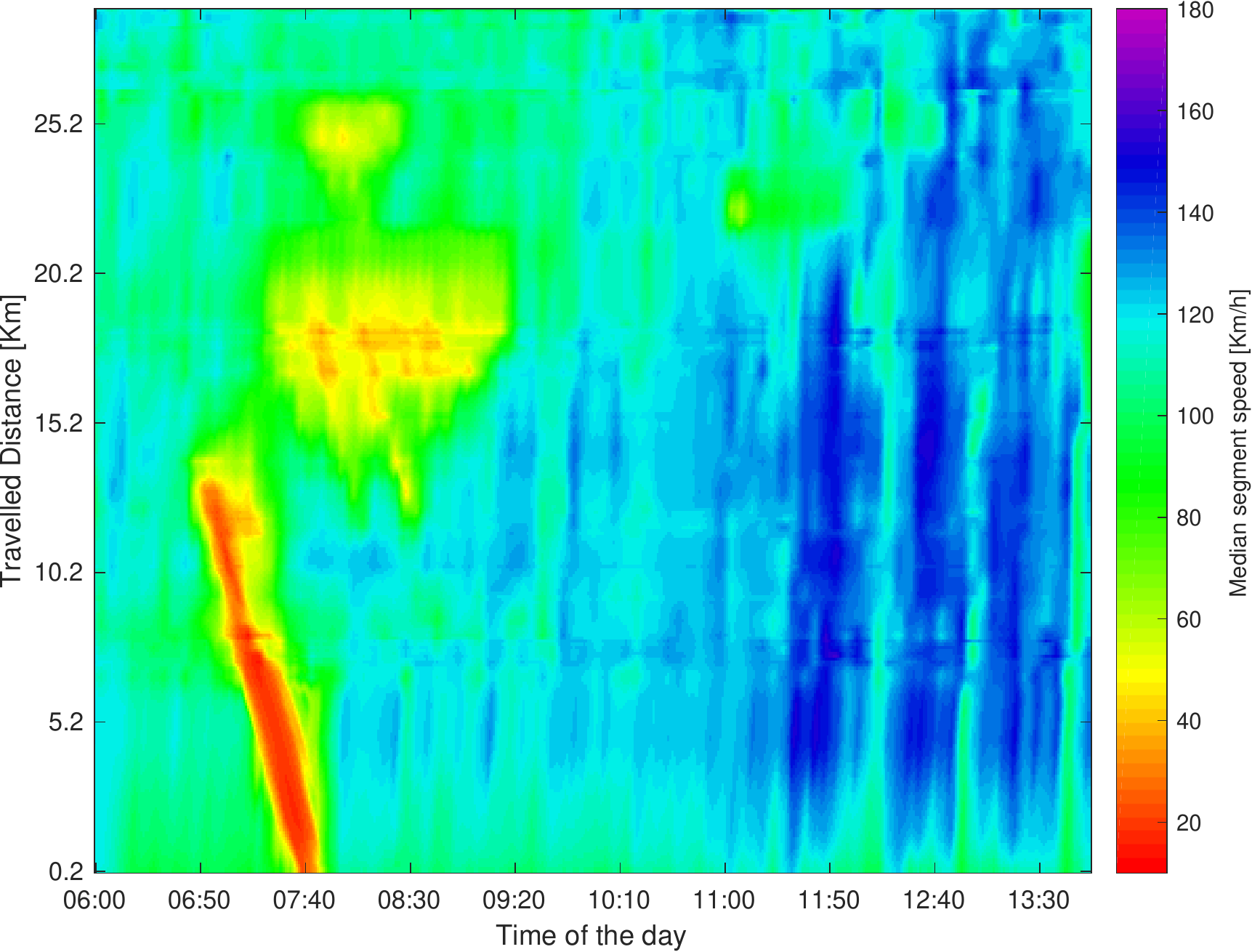}}
		\hspace*{\fill}%
	\subfloat[\label{monday4} \scriptsize{Monday, 22$^{nd}$ February}]{
		\includegraphics[width=0.45\textwidth]{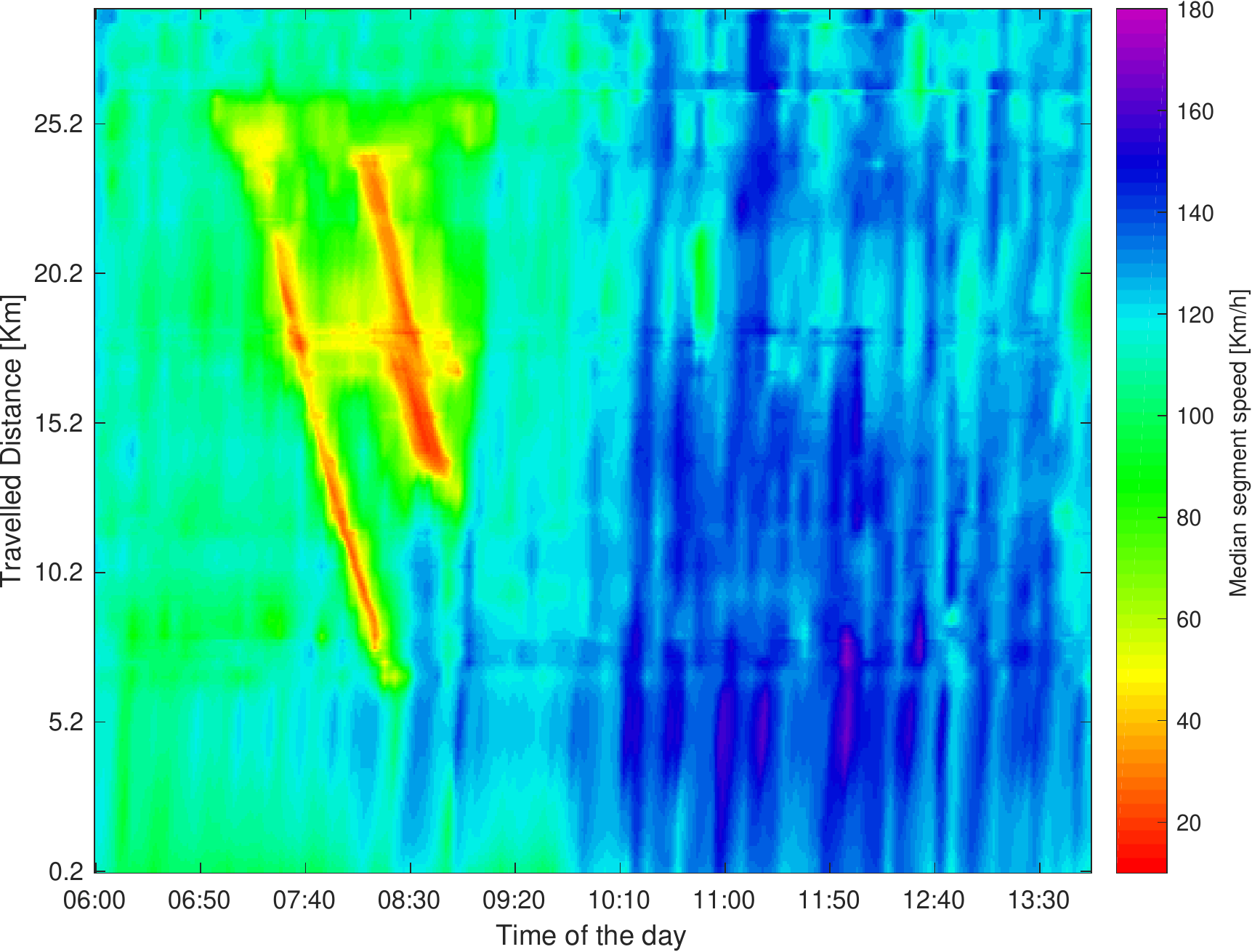}}\\
	\subfloat[\label{joint mondays}\scriptsize{Aggregated Mondays of February}]{
		\includegraphics[width=0.45\textwidth]{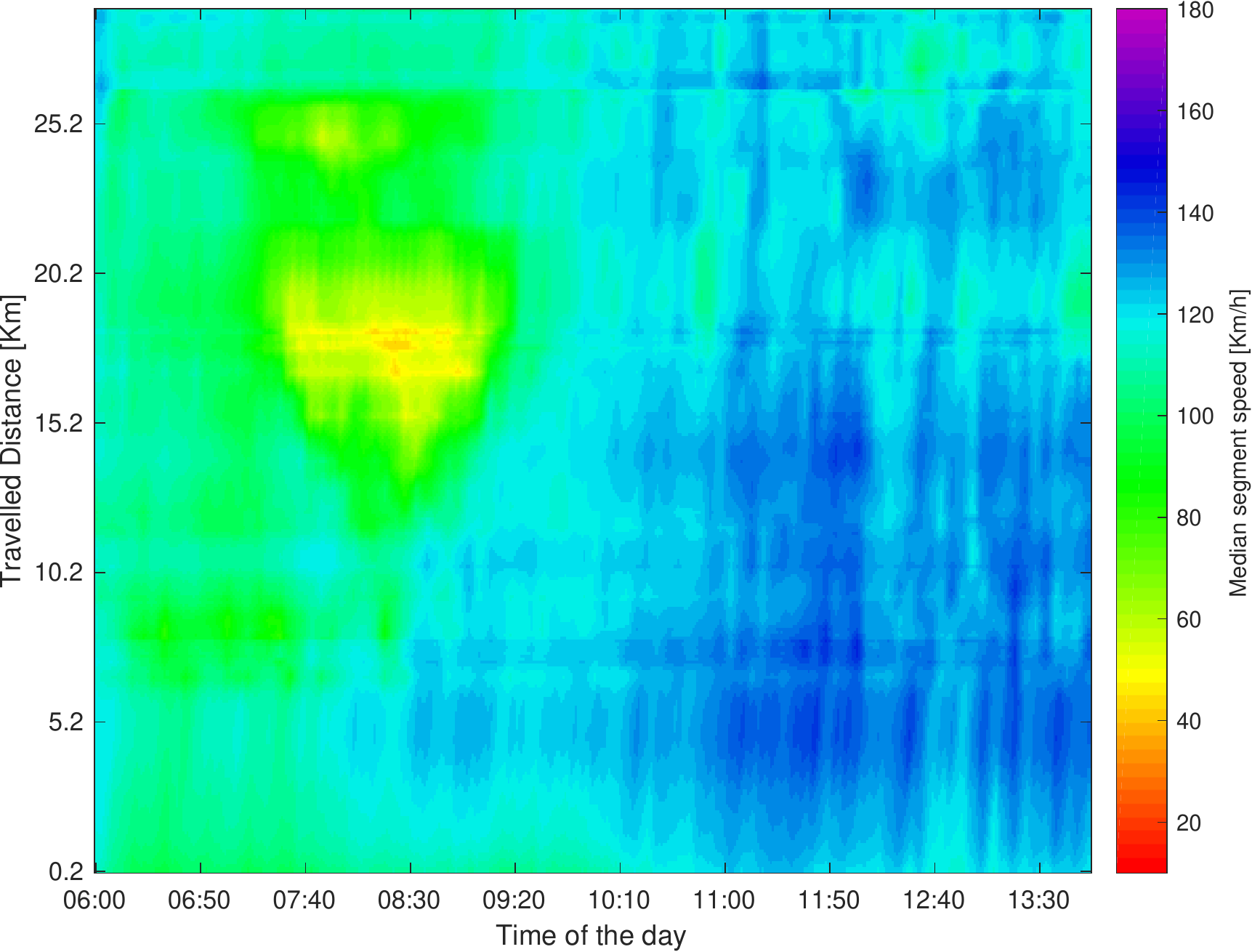}}
	\caption{ Comparison of speed fields for each Monday of February 2016 and an aggregated speed field for all Mondays of February 2016}\label{fig:aggregation effects}
\end{figure}   
 
In fig.~\ref{fig:aggregation effects} we can see the median speed field from a section of the German highway, A5-South (Km 465 to 492), near Frankfurt for each Monday of February 2016. Even though the capacity of this section has been greatly improved, by adding a forth lane in each direction \cite{extralane}, congestion still happens during morning rush hours as it can be seen in fig.~\ref{monday1}, \ref{monday3}, and \ref{monday4}. The speed field for the $8^{th}$ February, fig.~\ref{monday2}, shows no congestion because it is a national holiday in Germany. 
\begin{table}[t]
	\renewcommand{\arraystretch}{1}
	\centering
	\captionsetup{margin=1cm}
	\caption{ Parameters of the adaptive smoothing method with the values used in this experiment.}
	\label{asm parameters table}
	\scalebox{0.8}{
		\begin{tabular}{ll}
			\toprule
			\textbf{Parameter} & \textbf{Value} \\ \midrule
			Spatial smoothing width $\sigma $ &  $\Delta l_i/2$  \\
			Temporal smoothing width $\tau$ &  $\Delta t/2$ \\
			Propagation velocity of perturbations in free traffic $c_{free}$&  70 km/h\\ 
			Propagation velocity of perturbations in congested traffic $c_{cong}$&-15 km/h\\
			Threshold between free and congested traffic $V_c$& 50 km/h\\
			Width of the transition between free and congested traffic $\Delta V$&10 km/h\\ \bottomrule
	\end{tabular}}
\end{table}

\section{Time-dependence of travel times}\label{sec:timedependence}
\begin{figure}[tbh]
	\centering
	\captionsetup{margin=1cm}
	\includegraphics[width=0.75\textwidth, trim=0cm 0cm 0cm 0.35cm, clip]{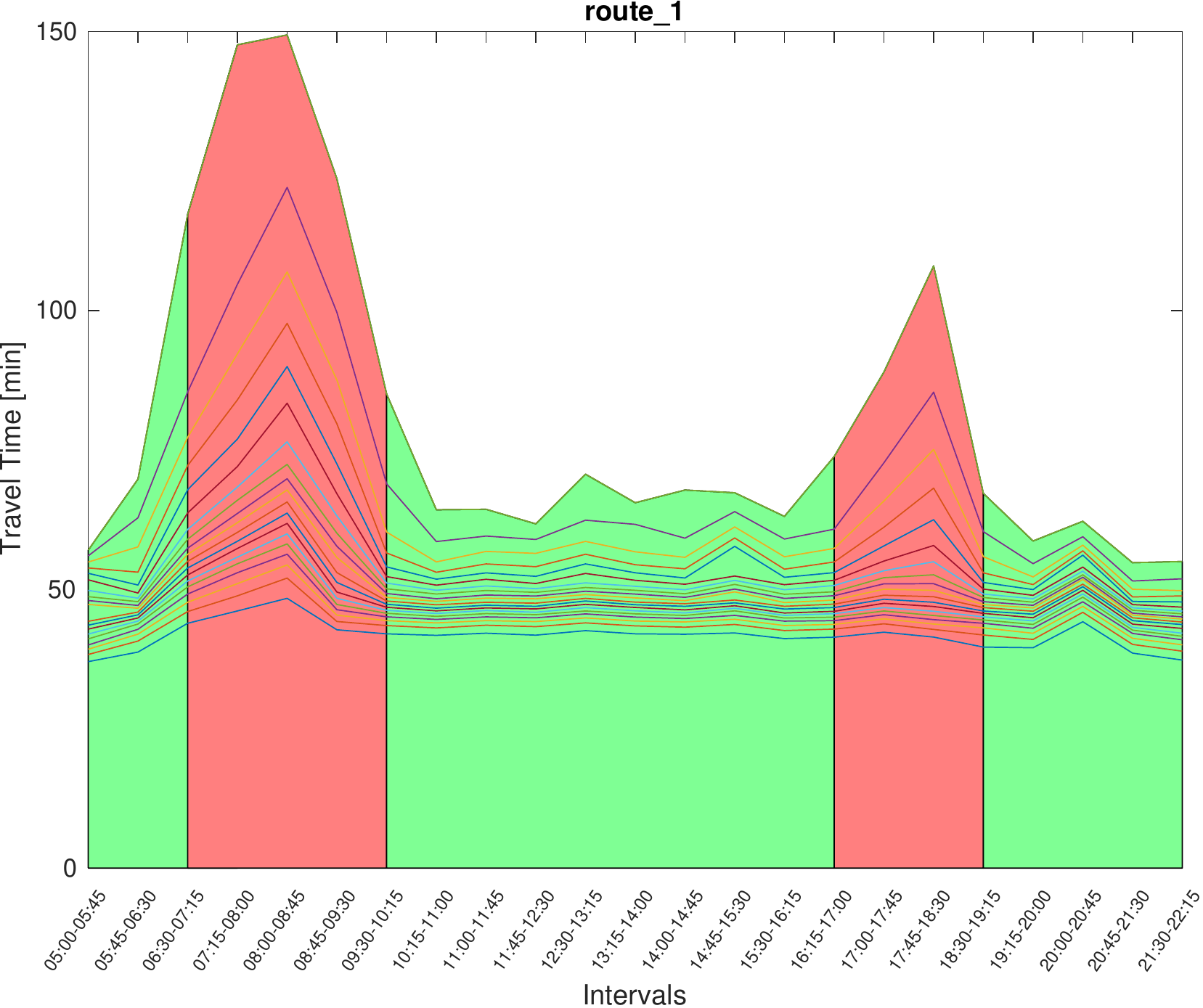}
	\caption{Morning and evening rush hours}\label{demandbasedintervals}
\end{figure}

Day-to-day fluctuations in travel time can range from a minor annoyance to a significant contributor of a traveller's decision-making process. Travel time reliability has become a key performance indicator of transportation networks and is mostly concerned with properties of the day-to-day travel time distributions \cite{van2008travel}. A number of key characteristics can be identified by analysing such distributions from empirical data. The wider (i.e. longer-tailed) it is on a particular time-of-day and day of the week, the more unreliable travel time will be for a given road stretch. The day-to-day distribution of travel times is a result of the fluctuations in both traffic demand and supply characteristics throughout the day. Since TomTom is constantly collecting data, we can analyse such fluctuations by narrowing (or widening) time-dependent variables, i.e. date range, day of week and time of day. Fig.~\ref{demandbasedintervals} for instance, shows the travel time percentiles $\left(5^{th}, 10^{th}, \ldots, 90^{th}, 95^{th}\right)$, from Zaandam to Enkhuizen on Tuesdays, in the date range of September 2015 to February 2016, as a function of the time of day. Both morning and afternoon rush hours are easily identifiable. Visualizations such as this one, give valuable insights into traffic for transportation planners. By identifying high demand periods of the day in specific routes, planners can avoid such time periods and decide for alternative trajectories. While using deterministic travel times, as most of the software packages do, such fluctuations are never taken into account.

\begin{figure}[tbh]
	\centering
	\captionsetup{margin=1cm}
	\includegraphics[width=0.75\textwidth, trim=0cm 0cm 0cm 0.35cm, clip]{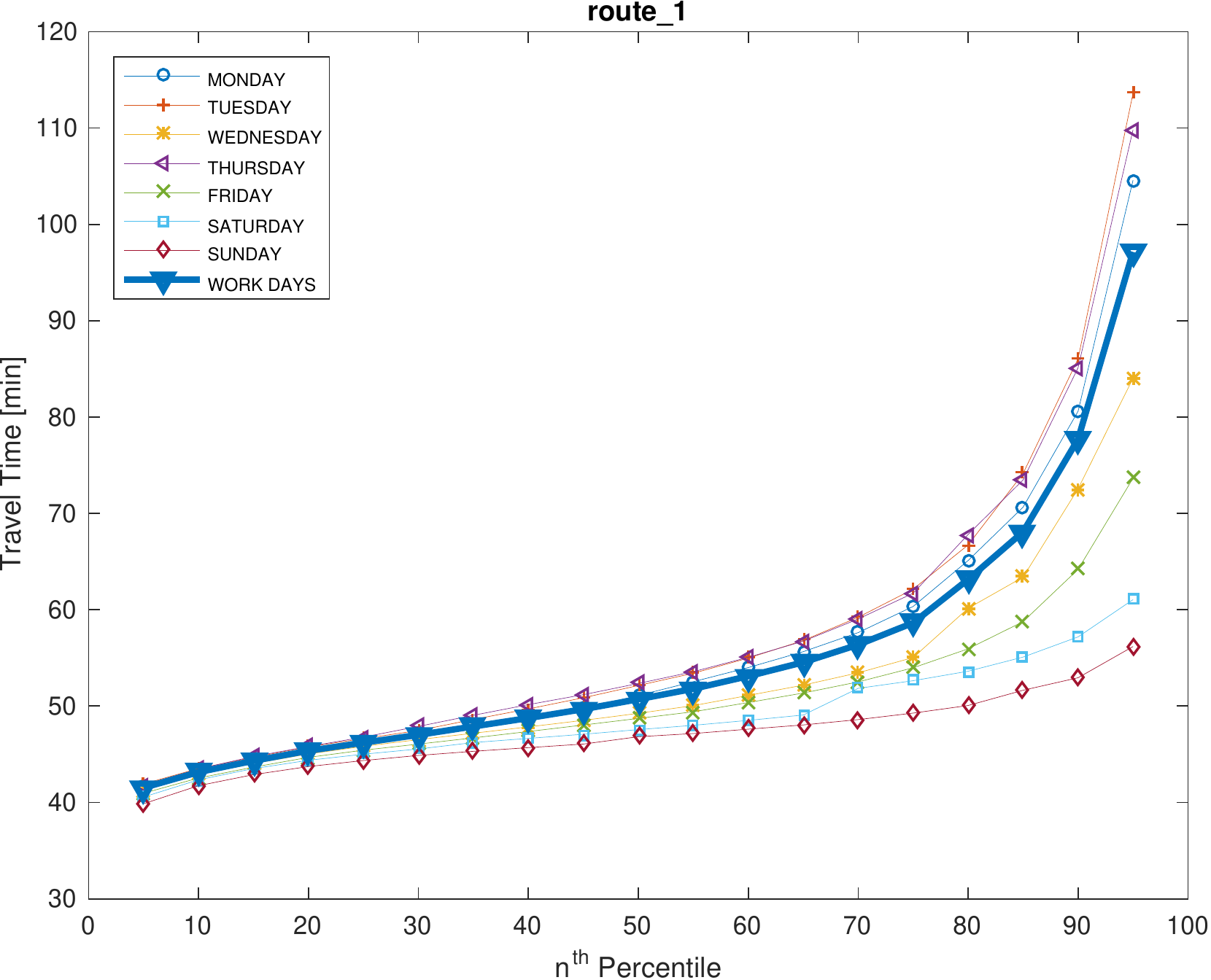}
	\caption{Travel time percentiles with collected from September 2015 to February 2016, between 08h00 and 09h00, for each day of the week and all working days.}\label{weekdays}
\end{figure}

\begin{figure}[tbh]
	\centering
	\captionsetup{margin=1cm}
	\includegraphics[width=0.75\textwidth, trim=0cm 0cm 0cm 0.35cm, clip]{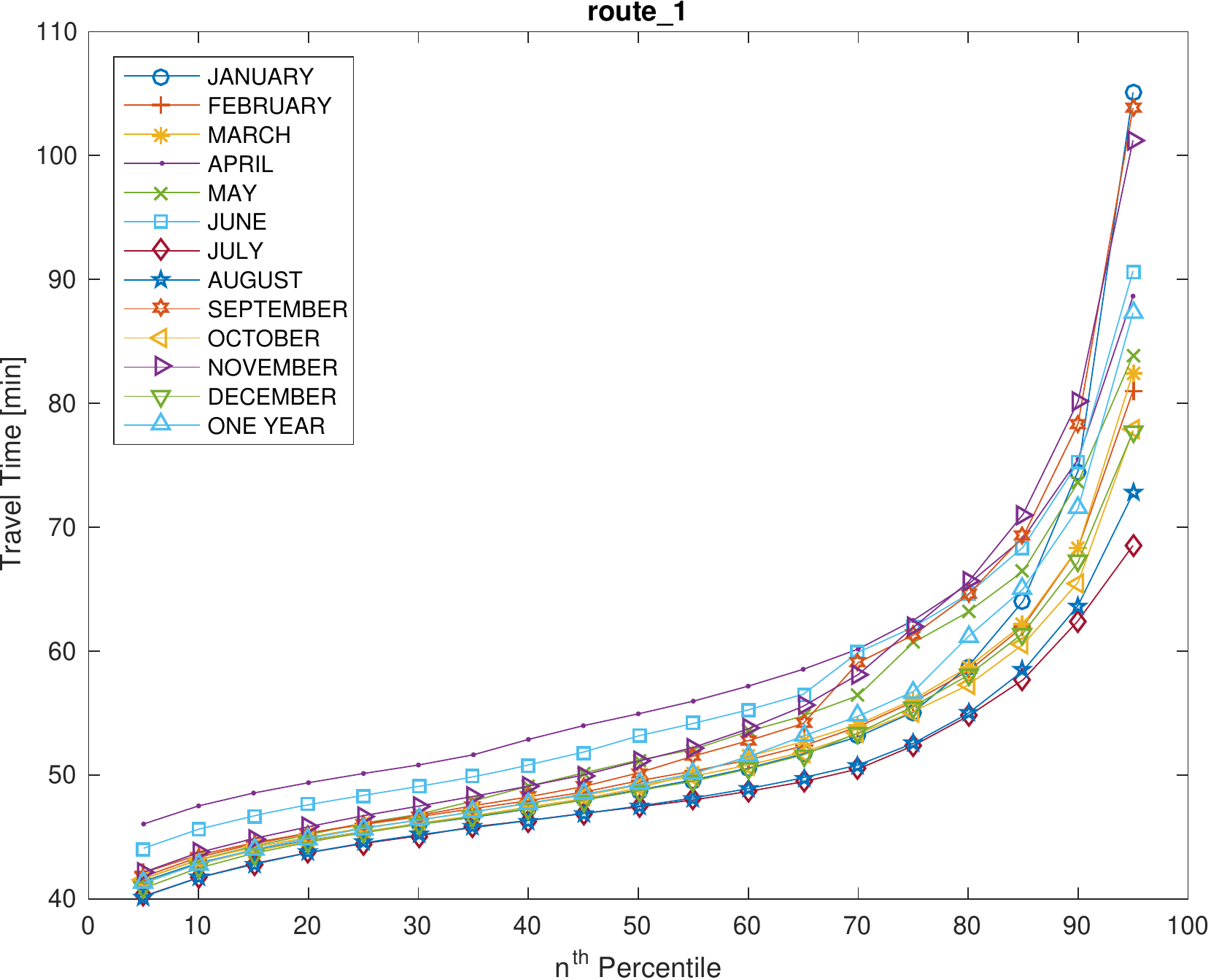}
	\caption{Travel time percentiles on Tuesdays, between 07h00 and 10h00, for each month of 2015 and for 2015 as a whole.}\label{months2015}
\end{figure}

Travel time changes are most notorious throughout the day, as seen in fig.~\ref{demandbasedintervals}. However, the day of the week can also have a significant impact in travel time. Fig.~\ref{weekdays} shows the percentiles of travel times throughout the week, during a rush-hour period, from 08h00 to 09h00. As expected, Saturday and Sunday are the fastest days to commute on this route. The busiest day is Tuesday, closely followed by Thursday and Monday, respectively. What is particular interesting to observe is that, for the 5\% \textit{luckiest} drivers differences in travel time on different days of the week were of only a couple of minutes, while for the 5\% most \textit{unfortunate} drivers differences were huge. Between Sunday and Tuesday, for instance, the difference is of 55 minutes. Considering the uncongested travel time of around 45 minutes, that is an increase of 122\%. This results however, are only valid for the selected date range.
Fig.~\ref{months2015} shows the percentiles of travel times also on Tuesdays, from 07h00 to 10h00, for 13 different date ranges, that is one for each month of 2015 and one for the complete year. The difference between work-related and holiday-related months is huge. January, September and November are the busier months of the year. 95\% of the travels, for this route, take up to 105 minutes in these months. In contrast, during July, the same route takes only up to 70 minutes, 95\% of the times. As expected, when using the one year range, travel times percentiles are averaged out.

\section{Estimating travel times distributions}\label{sec:estimate}
\begin{figure}[tbh]
	\centering
	\captionsetup{margin=1cm}
	\subfloat[\label{roadlinks}]{
		\includegraphics[width=0.4\textwidth]{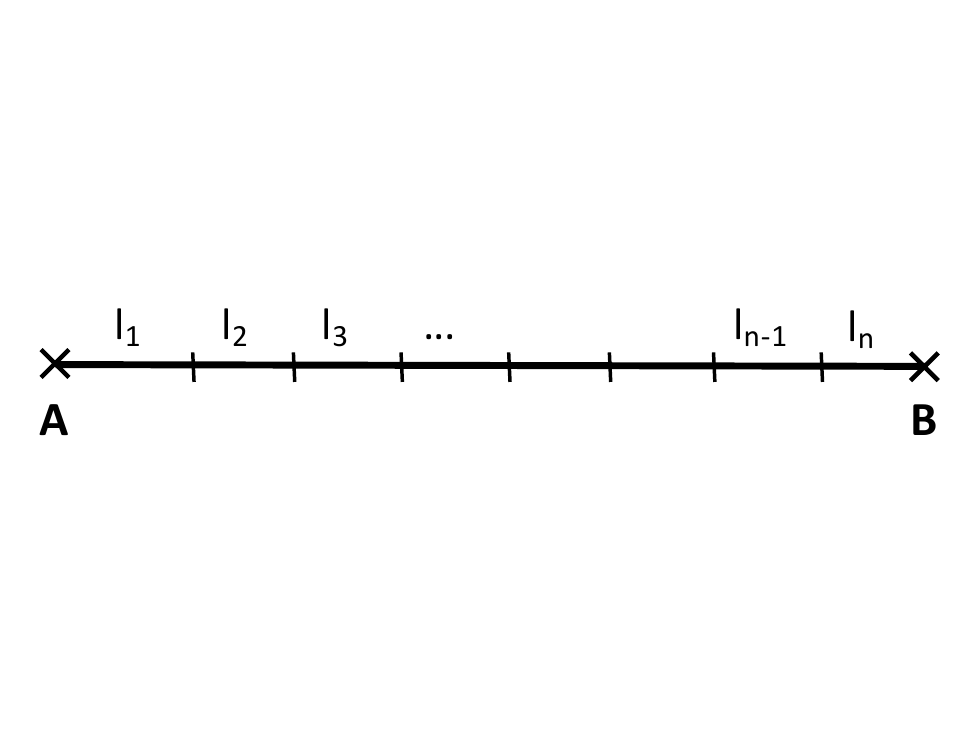}}
	\hspace{\fill}
	\subfloat[\label{linkdist}]{
		\includegraphics[width=0.45\textwidth]{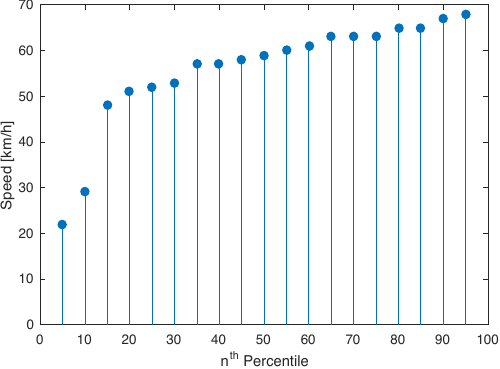}}
	\caption{(a) Road links for route $AB$; (b) Speed distribution for a link $l_i$}\label{fig:traject}
\end{figure}

The way in which travel times distributions can be aggregated from TomTom data changes, depending on how the time-dependent variables are defined. The aggregation frame, i.e. how data is partitioned in space and time, has a big influence on the estimated travel times mostly due to the time-dependencies described above. In space, there are two options to consider for the aggregation frame: either route-based or link-based. Consider a route from $A$ to $B$, composed by $n$ links (see fig.~\ref{roadlinks}):  (i) in a route-based space aggregation frame, travel time (and speed) is described by a single cumulative distribution with respect to the complete trajectory; (ii) in a link-based space aggregation frame, there will be $n$ cumulative speed distributions, i.e. one for each link, see fig.~\ref{linkdist}. Since the length of each link is known, link travel time distributions can be easily obtained.
The aggregation frame in time, can be freely selected and is defined by the \textit{time of day} intervals. Shorter \textit{TOD} intervals, in the minutes order of magnitude, yield time-specific estimations but they can lack in accuracy due to smaller sample sizes. On the other hand, larger intervals tend to be more accurate due to the higher number of probe vehicles included on measurements but they are less precise given their wider time-coverage. Depending on how \textit{TOD} intervals are defined, travel times are computed differently. For this experiment we evaluated a set of routes using three different sets of \textit{TOD} intervals. A 24h day can be divided into: (i) 5 minute intervals; (ii) 20 minute intervals or (iii) 5 custom intervals based daily traffic demand levels, see fig.~\ref{demandbasedintervals}.

For short intervals, i.e. shorter than the expected duration of the trip, as in (i) and (ii), travel times must be computed in a link-based manner, by summing all link travel times at the moment the vehicle enters the link. In situations where a given road link has no data available, the local speed is considered equal to the speed limit of the link. The link speed distribution becomes a constant. The assumption is that if there are no vehicles equipped with a TomTom device traversing a link on a given period, the link is in free flow and can be travelled at the maximum allowed speed.

 If we were to estimate travel times by sampling the overall route-based distribution, the estimation would be either too high or to low, depending on the characteristic traffic demand of that short period. For instance, consider a short 5 minute interval from 08h55 to 09h00. Since this is commonly a high traffic period (i.e. morning rush-hour), during this interval, all road links will show relatively slow speeds compared with other periods. The travelling vehicle, however, will only experience such slow speeds in some of the road links, not during the whole length of the trip, making the estimated value to be too high. 

When considering larger intervals, as in (iii), travel time can be estimated just by sampling the overall route-based distribution because it already takes into account the travel time fluctuations that took place throughout the entire trip. When considering short \textit{TOD} intervals, travel time has to be estimated iteratively, by summing all link travel times at the moment that the vehicle enters the link.
\begin{figure}[tbh]
	\centering
	\captionsetup{margin=1cm}
	\includegraphics[width=0.45\textwidth]{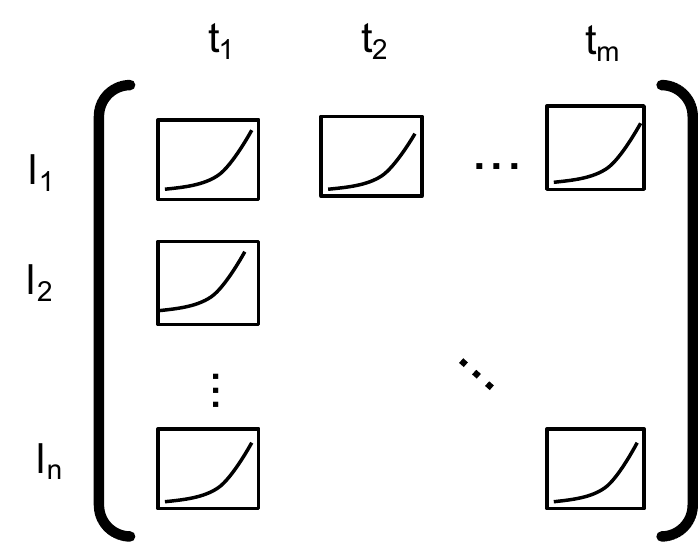}
	\caption{Link travel time matrix: each entry is a travel time distribution for link $l_i$  during \textit{TOD} interval $t_j$.}
\end{figure}
Link travel time dependencies are another aspect to consider when estimating travel times from link-based speed distributions. The naive method is to assume independence of link travel times (i.e. speeds), randomly sample link speed distributions and consequently sum the individual link travel times along the route at the moment the vehicle enters the link. In this approach, correlation between links is assumed as zero and travel time calculation is straightforward. However, assuming independence is not realistic. During a peak period it would be expected for link travel times to increase for all links constituting a route. However, if the travel time on a link increases due to congestion, it would also be expected the downstream links to have lower travel times, while upstream links would have higher travel times. Since we know the speed distributions for each link on a given TOD, we can model driver's behaviour by constraining how the vehicle speed changes from link to link. Instead of randomly sampling all link speed distributions, we only do it for the first link. That first draw from the distribution, sets initial vehicle speed. Consequent speeds are sampled from the following link speed distributions constrained by $\alpha$, the maximum acceleration rate from link $l_{i-1}$ to link $l_i$. In other words, link speed distribution of link $l_i$ is limited to the range defined by the previous sampling value of $l_{i-1}$ $\pm$ $\alpha$.
\begin{figure}[tbh]
	\centering
		\captionsetup{margin=1cm}
	\subfloat[\label{alpha=1} $\alpha=1$]{
		\includegraphics[width=0.95\textwidth, trim= 2.4cm 0.25cm 2.4cm 0.7cm,clip]{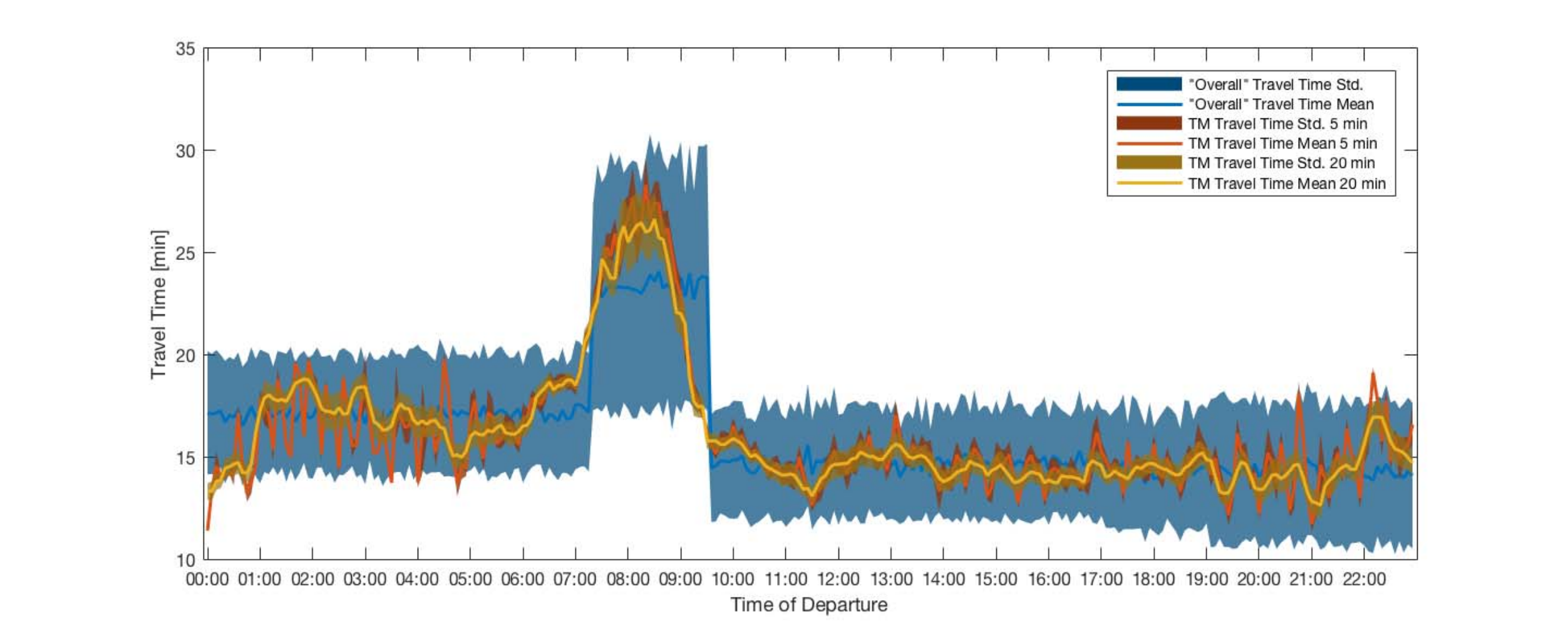}}\\
	\subfloat[\label{alpha=0.1}$\alpha=0.1$]{
		\includegraphics[width=0.95\textwidth, trim= 2.4cm 0.25cm 2.4cm 0.7cm,clip]{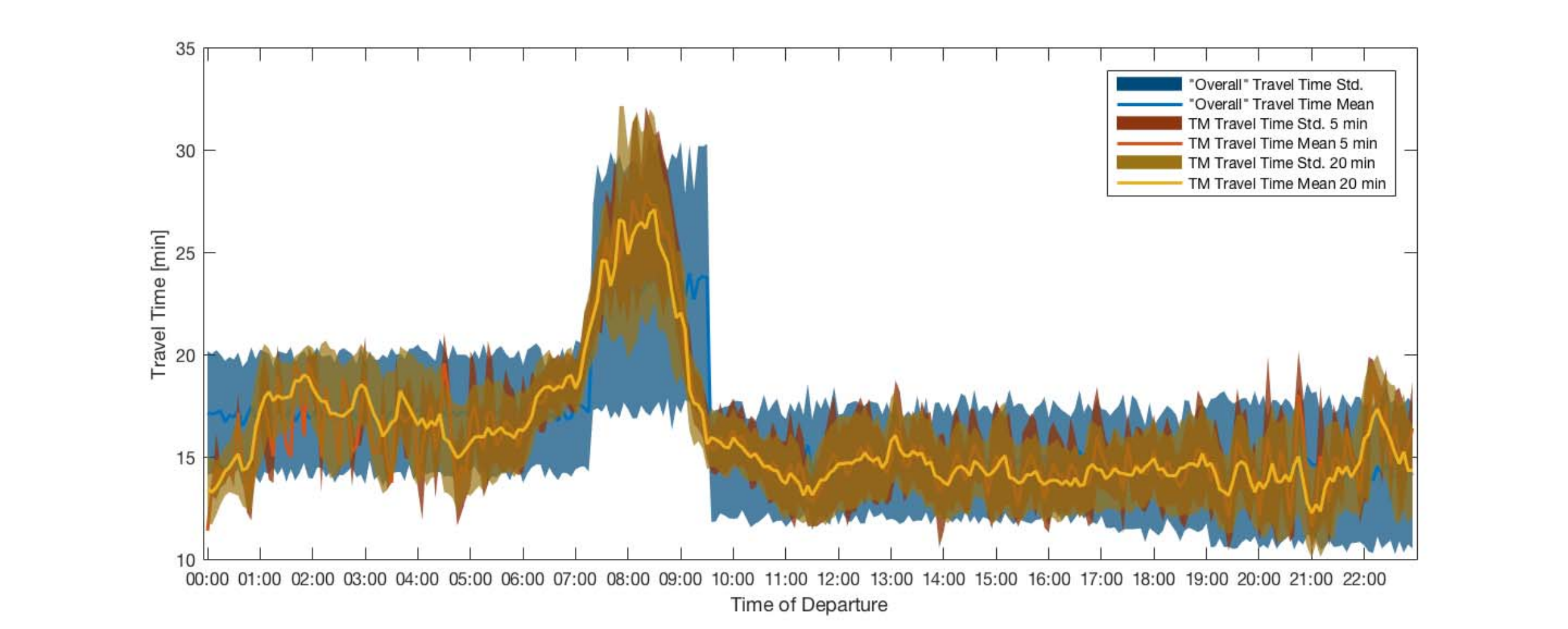}}
	\caption{Travel time estimation with and without dependences between link travel times}\label{fig:alphas}
\end{figure}

Fig.~\ref{fig:alphas} shows the estimated mean travel times $\pm$ standard deviation as function of the departure time. For each trip, departure time and type of TOD intervals, 500 simulations were made. The data being shown was collected on Monday, $1^{st}$ of February, 2016 from 00h00 to 23h00. Each colour refers to the type of TOD intervals used in the estimation process: 5 min intervals, 20 min intervals or 5 larger intervals as in fig.~\ref{demandbasedintervals}. Looking at the estimation made by route-based distributions, the blue line, we can distinguish three zones: early morning, morning rush-hour and the rest of the day. In each one, mean travel time is kept fairly constant ($\pm 18 \rightarrow 24 \rightarrow 15$ min, respectively) and the same goes for the standard deviation ($\pm 3 \rightarrow 5 \rightarrow 2.5$ min, respectively). These are very stable values because it is the same distribution being sampled over and over again. Since these distributions cover a wide time frame (e.g. hours) they show high standard deviations when compared to the 5 min and 20 min TOD estimations that are almost negligible in fig.~\ref{alpha=1}. Apparently, on this route, afternoon rush-hour does not manifest and therefore we cannot see the 5 distributions used in the blue estimation. The orange and yellow lines represent the estimated mean travel time for the 5 and 20 min TOD intervals, respectively. Since these are computed on a link by link basis, estimations are much more dependent on the departure time. Instead of having 5 distributions to describe a 24h time-span, there are 288 (72) for 5 (20) min TOD intervals. Because link speed distributions are sampled randomly (i.e. when $\alpha=1$), speed fluctuations from link to link eventually cancel each other and the estimated travel time is always close to the mean value, especially for the 5 min TOD intervals. Since these are very short periods, the amount of vehicles traversing these links is also small which leads to very narrow link speed distributions. In fig.~\ref{alpha=0.1}, standard deviations are much bigger for both cases with 5 and 20 min TOD intervals because the assumption of independent link speeds was dropped.

\section{Predicting travel times distributions}\label{sec:prediction}

The previous section shows how travel time distributions can be accurately generated based on historical data. In order to create use that information to create a transportation plan, historical data must be selected from which a travel time distribution can be generated that accurately predicts the future travel time. For example, if we want to predict the travel time distribution for a particular route for tomorrow and tomorrow is a Monday, we can generate a travel time distribution based on all data from the past week, all data from last Monday, all data from the last four Mondays, etc.

This illustrates that different variables apply when selecting historical data to generate a travel time prediction from. In particular, we identified the variables:
\begin{compactitem}
\item number of historical days;
\item level of granularity; and
\item assumed speed relation between segments.
\end{compactitem}
To predict travel times for a particular weekday, we only historical select data from the same weekday, because descriptive analytics clearly shows that different days of the week lead to different travel times. However, this still leaves a decision on the number of days to use data from.

\begin{figure}[tbh]
	\centering
	\captionsetup{margin=1cm}
	\subfloat[\label{reference1} \scriptsize{Estimation - Reference, 29$^{nd}$ February}]{
		\includegraphics[width=0.5\textwidth]{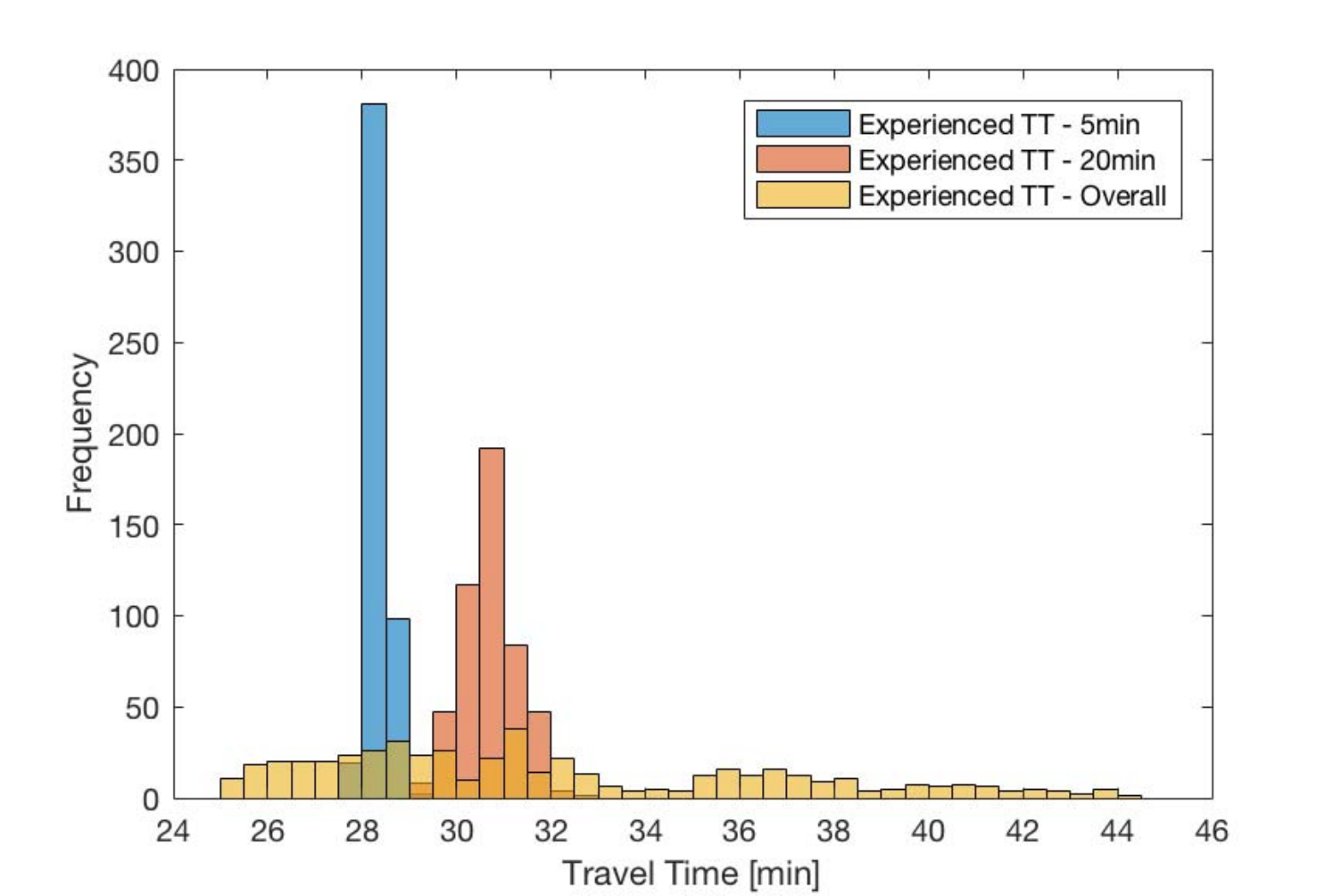}}
		\hspace*{\fill}%
	\subfloat[\label{prevweek1} \scriptsize Prev. Week, 22$^{th}$ February]{
		\includegraphics[width=0.5\textwidth]{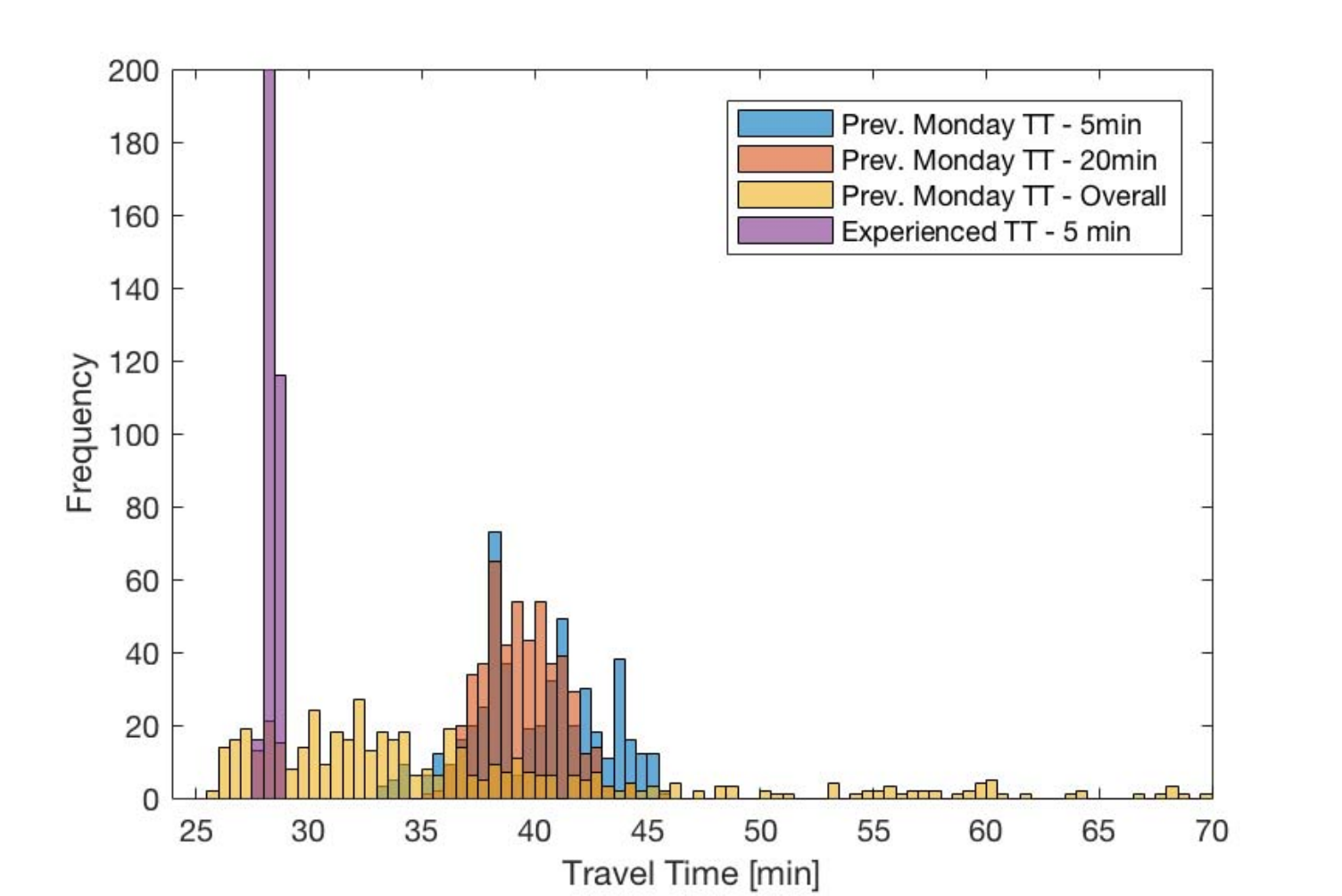}}\\
	\subfloat[\label{prevmonth1} \scriptsize{Prev. Month}]{
		\includegraphics[width=0.5\textwidth]{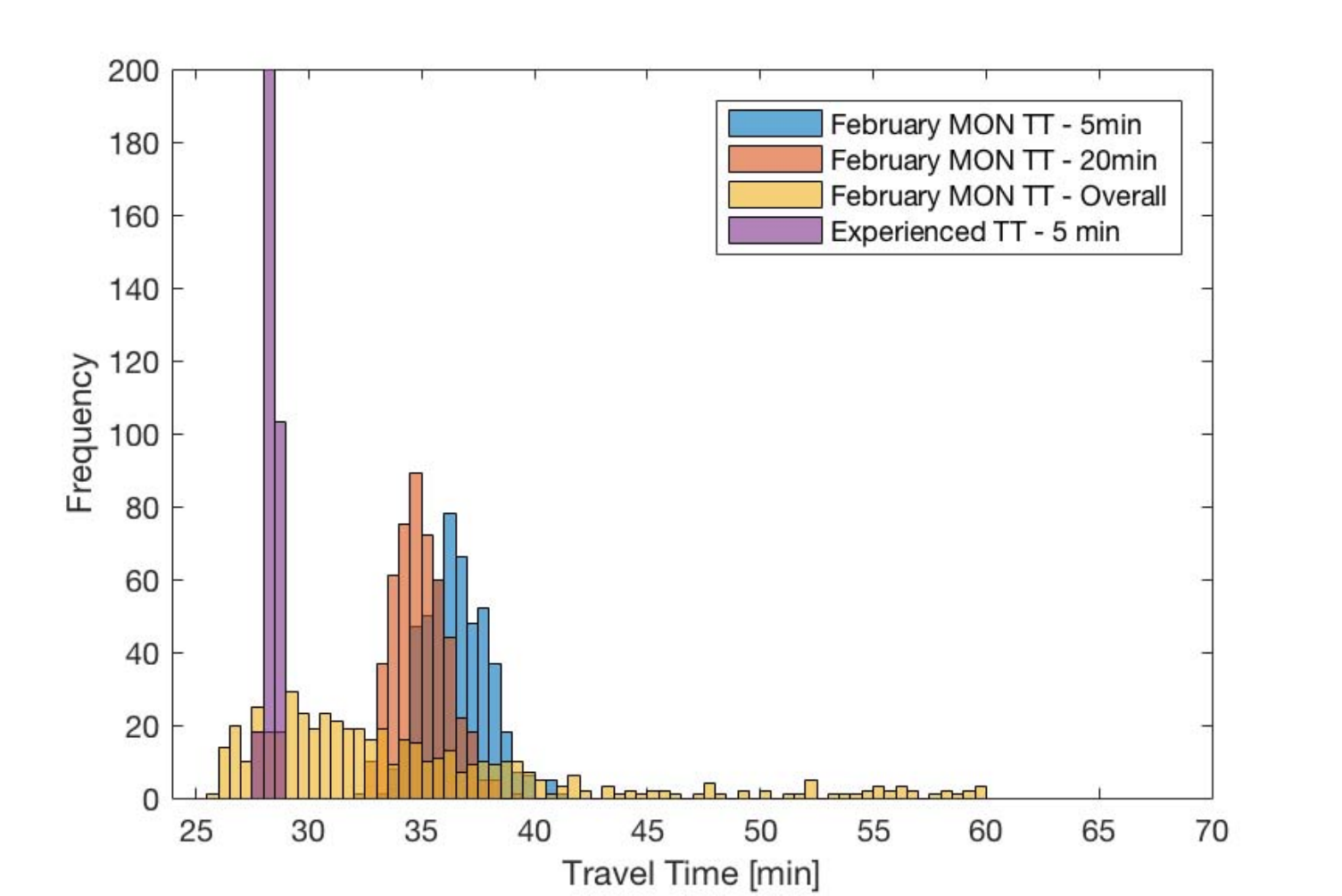}}
		\hspace*{\fill}%
	\subfloat[\label{prev3month1} \scriptsize{Prev. 3 Months}]{
		\includegraphics[width=0.5\textwidth]{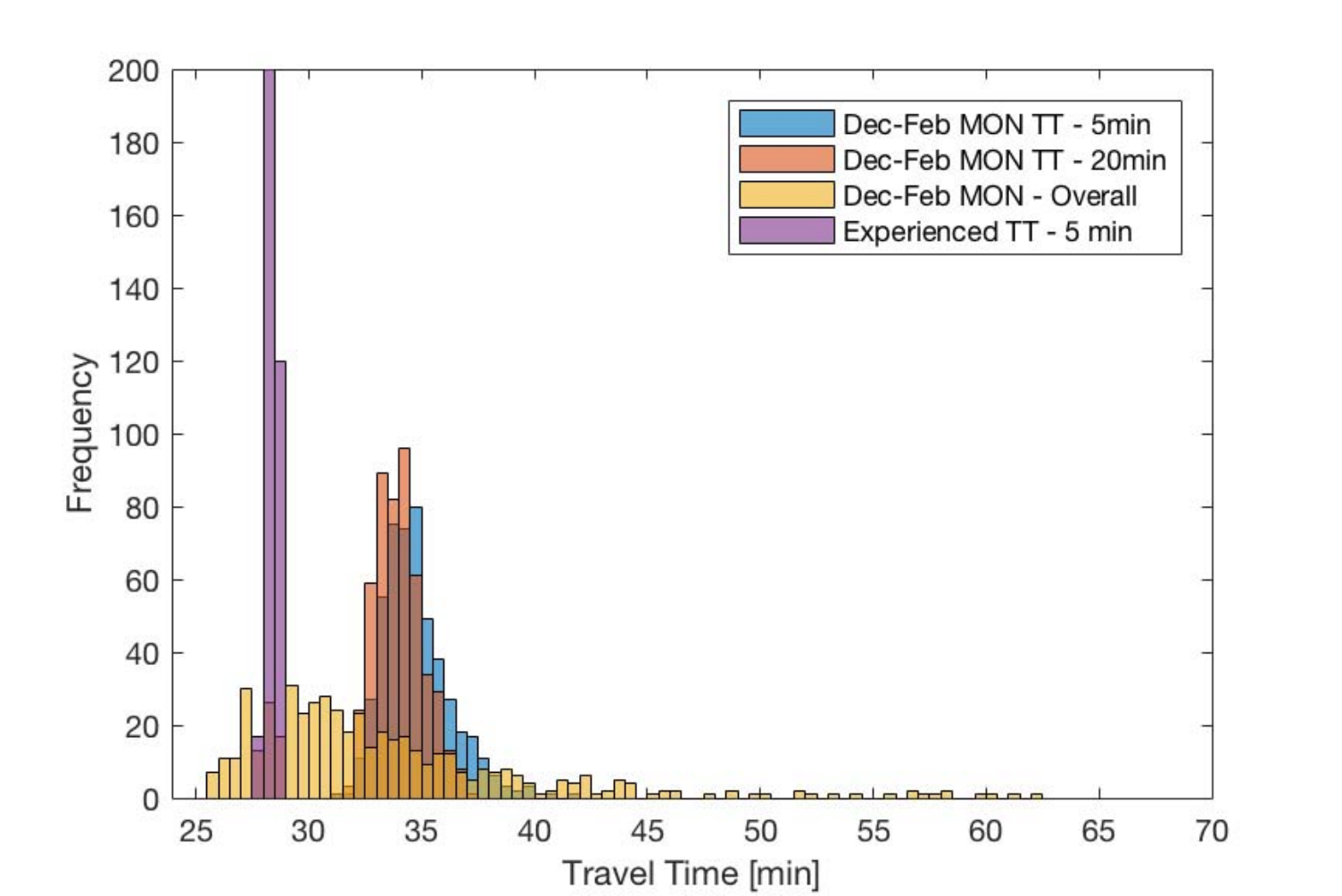}}
	\caption{$\alpha=1$}\label{forecastalpha1}
\end{figure}

\begin{figure}[tbh]
	\centering
	\captionsetup{margin=1cm}
	\subfloat[\label{reference01} \scriptsize{Estimation - Reference, 29$^{nd}$ February}]{
		\includegraphics[width=0.5\textwidth]{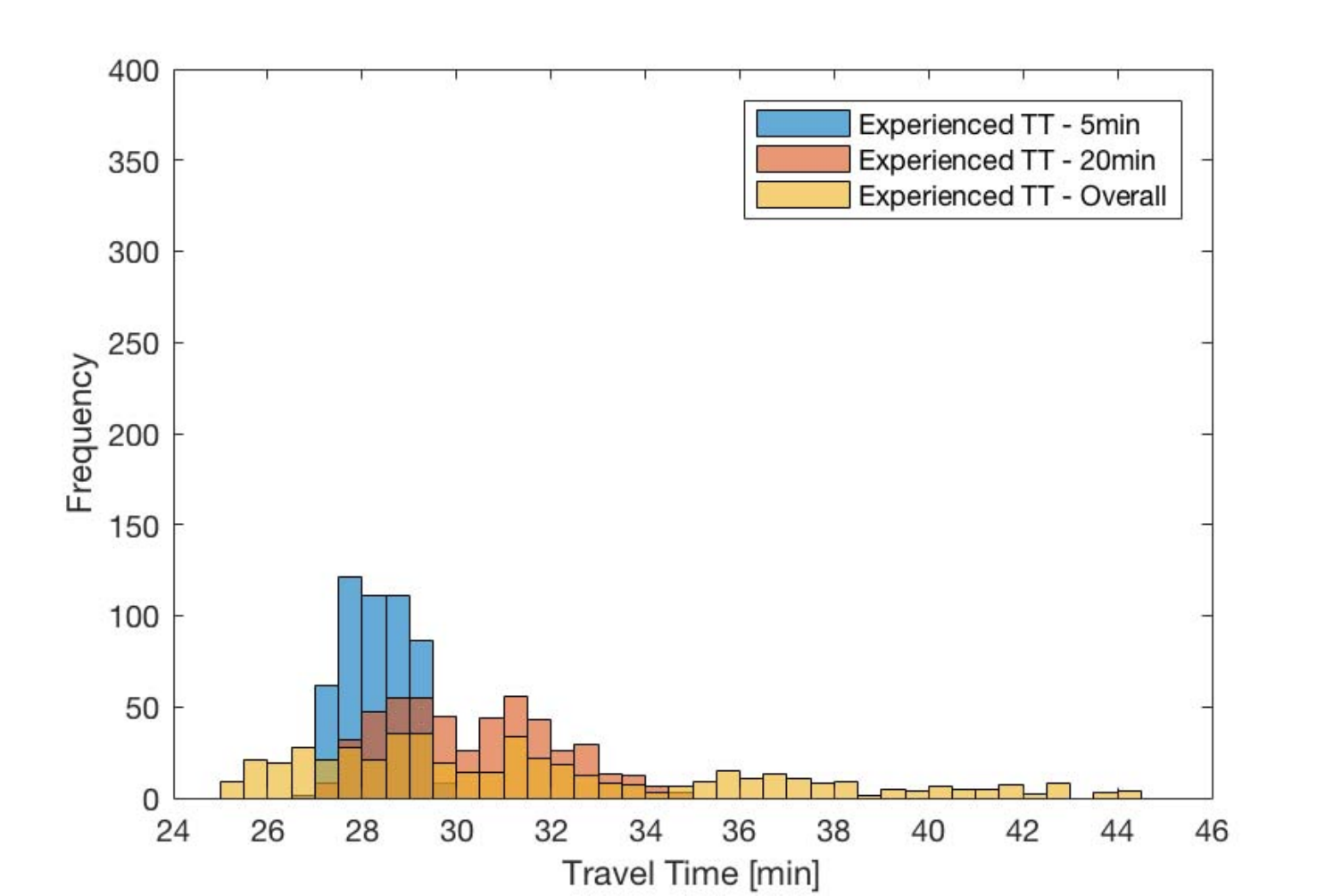}}
			\hspace*{\fill}%
	\subfloat[\label{prevweek01} \scriptsize Prev. Week, 22$^{th}$ February]{
		\includegraphics[width=0.5\textwidth]{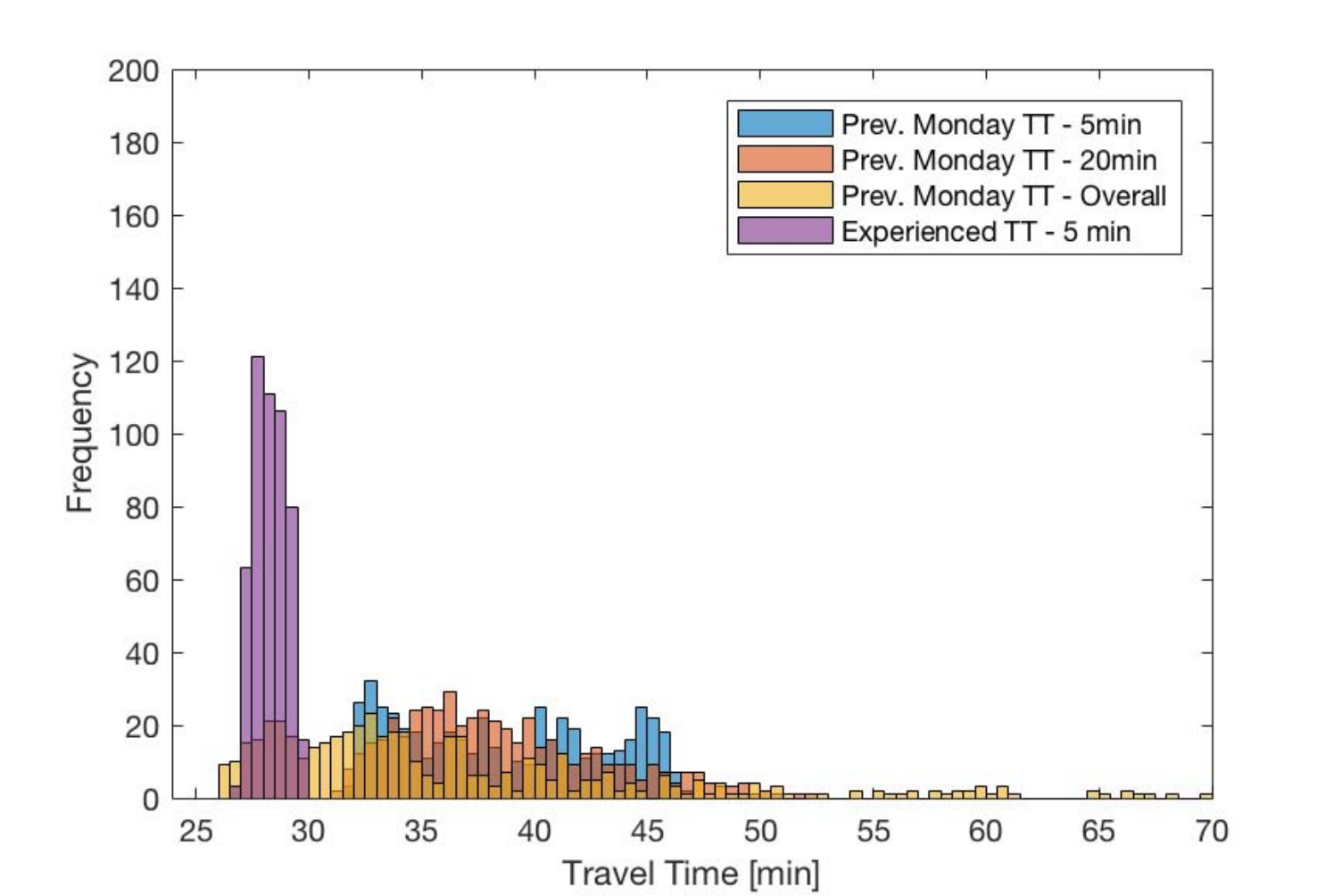}}\\
	\subfloat[\label{prevmonth01} \scriptsize{Prev. Month}]{
		\includegraphics[width=0.5\textwidth]{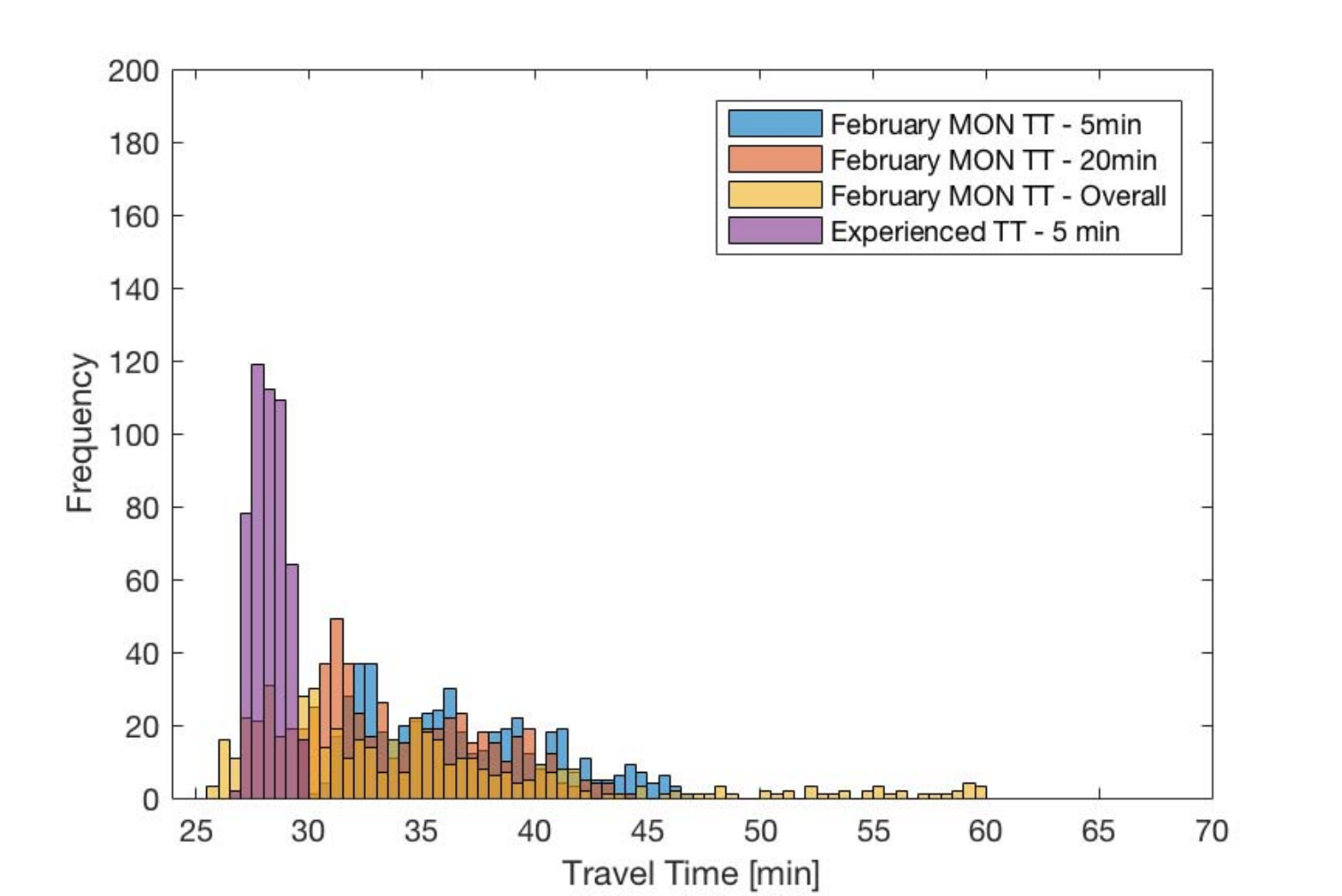}}
			\hspace*{\fill}%
	\subfloat[\label{prev3month01} \scriptsize{Prev. 3 Months}]{
		\includegraphics[width=0.5\textwidth]{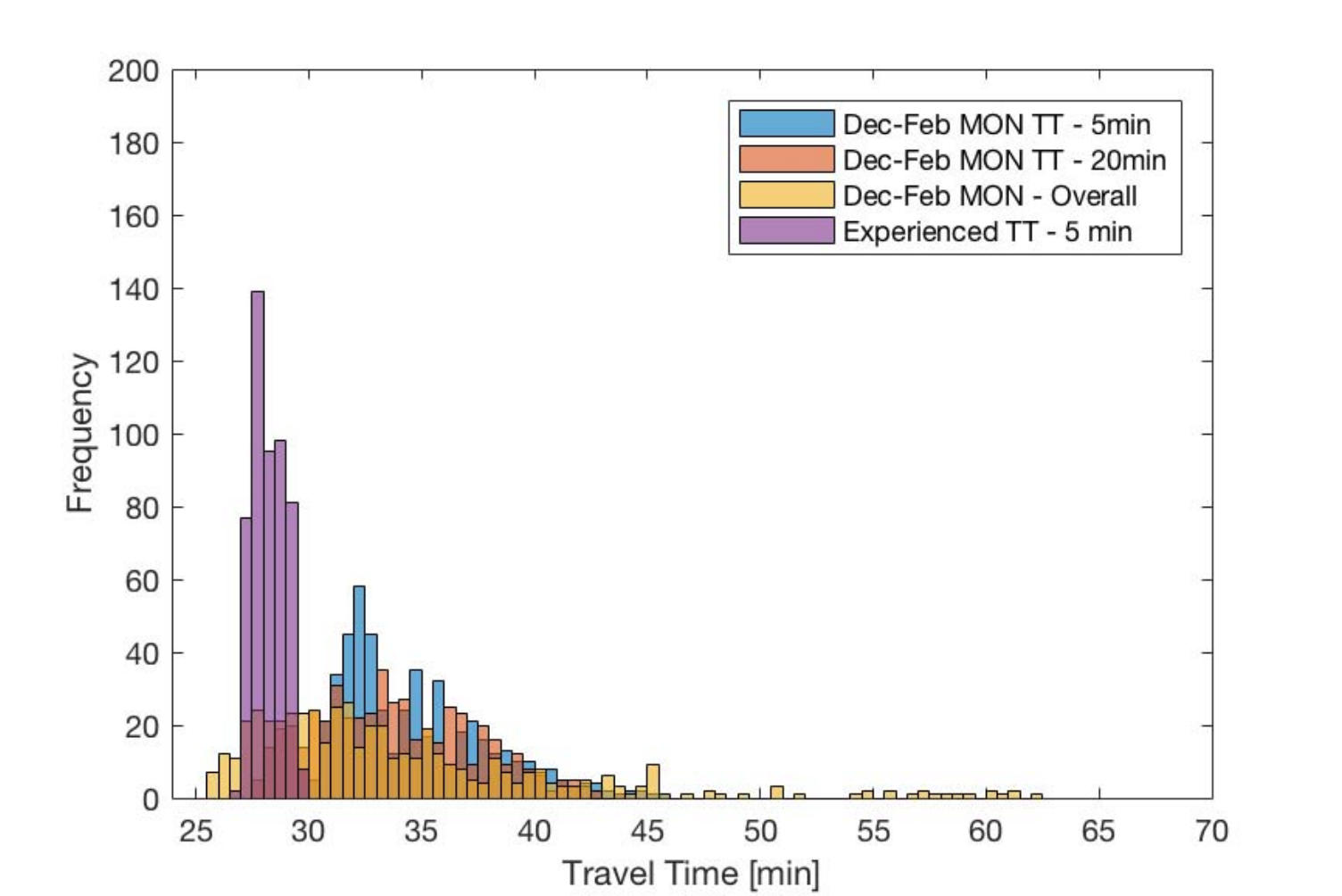}}
	\caption{ $\alpha=0.1$}\label{forecastalpha0.1}
\end{figure}

To illustrate the effect of these variables, we compare the travel time predictions for the $29^{th}$ of February 2016 leaving at 8:30 for different values of these variables and for the route that we used in the previous section. Figure~\ref{forecastalpha1} and~\ref{forecastalpha0.1} show the results of these predictions. These figures show results based on an assumed speed relation between segments of $\alpha = 1$ (Figure~\ref{forecastalpha1}) and $\alpha=0.1$ (Figure~\ref{forecastalpha0.1}). In both these figures we used 1 historical day, 4 historical days (all days from the last month), and 12 historical days (all days from the last three months). We used the levels of granularity that we also used in the previous section: 5 minutes, 20 minutes, or 5 intervals during the day.

In both figures, sub-figure (a) shows the actual travel time estimate on $29^{th}$ of February, which is computed as explained in the previous section. The most accurate travel time estimate (at a 5 minute level of granularity) shows that the actual travel time on the route at 8:30 is estimated to be a little more than 28 minutes, with a rather low variance at $\alpha = 1$ and a slightly higher variance at $\alpha = 0.1$. The other sub-figures compare this estimate to travel time distributions that are predicted with the various settings.

The figures show that, as expected, a higher level of granularity and a higher $\alpha$ produce a more narrow travel time distribution. However, in all cases the estimate is unlikely or even impossible according to the predicted travel time distributions at the higher levels of granularity (5 minutes and 20 minutes). The predicted travel time distributions would on-average produce a value that is several minutes off. Predicted travel time distributions at the level of granularity of five intervals per day produce better results. The reason for these results is that the exact moment at which traffic congestion occurs is hard to predict and varies largely from one day to another, as was illustrated in Figure~ref{fig:aggregation effects}. While high-granular predictions predict this moment at 5 or 20 minute precision, our results show that this level of precision cannot be achieved. It must be noted that these results are different for quiet times of the day, when the chance of encountering congested traffic is low anyway. However, for those times of the day, higher granularity provides less added value, because for those times, travel times are less variable and therefore the low granularity predictions also provide accurate estimates. The setting of $\alpha$ does not make a large difference, although the results at $\alpha = 1$ appear to be slightly better. Table~\ref{od pairs} shows similar results, but at departure time 9:00, and only for $\alpha = 1$. At 9:00 the higher levels of granularity perform better, with the estimate (of 31.63 minutes travel time) less than one standard deviation away from the forecast at a 20 minute level of granularity using 1 or 4 days of historical data.

\begin{table}[tbh]
	\renewcommand{\arraystretch}{1}
	\centering
	\caption{ {Travel time forecasts, in minutes, with departure time 09h00 and $\alpha=1$}}
	\label{od pairs}
	\scalebox{0.74}{
		\begin{tabular}{lccccc}
			\toprule
			&\textbf{Range}&\textbf{TOD}&\textbf{Mean TT}&\textbf{TT Std.}&\textbf{Sample Size}\\\midrule
			\multirow{3}{*}{\textbf{Estimation - Reference}}&\multirow{3}{*}{1 Day}&5 min&31.63&0.20&2 \\
			&&20 min&31.55&0.51&5\\    
			&&Overall&31.91&4.67&50\\\midrule
			\multirow{3}{*}{\textbf{Forecast - Prev. Week}}&\multirow{3}{*}{1 Day}&5 min&29.22&0.79&2 \\
			&&20 min&31.66&0.73&5\\    
			&&Overall&37.16&10.56&50\\\midrule
			\multirow{3}{*}{\textbf{Forecast - Prev. Month}}&\multirow{3}{*}{4 Days}&5 min&30.92&0.38&5 \\
			&& 20 min&32.05&0.74&20\\    
			&&Overall&34.53&7.37&200\\\midrule
			\multirow{3}{*}{\textbf{Forecast - Prev. 3 Months}}&\multirow{3}{*}{12 Days}&5 min&32.74&1.04&20 \\
			&&20 min&33.12&1.39&20\\    
			&& Overall&34.19&7.21&500\\\bottomrule
	\end{tabular}}
\end{table}

From these results, we conclude that making travel time predictions is best done at a low level of granularity (i.e. collecting data in 5 intervals: morning rush hour, daytime, evening rush hour, evening, and night). While a higher levels of granularity produces a more precise predicted travel time distribution (i.e. with a lower standard deviation), the probability that the actual travel time is produced by that travel time distribution is lower, and often much lower, especially during times or on routes with high traffic volumes. Travel time predictions are also best done based on a relatively low number of recent historical days that have the same weekday as the day that must be predicted, because travel times vary a lot throughout the year. Public holidays and school vacations must be carefully considered when selecting historical days to base travel time predictions on.

\section{Conclusions}\label{sec:conclusions}
In this paper we showed how time-dependent stochastic travel times can be calculated for a route based on historical high-volume floating car data. Since only a few cars can be expected to have driven the same route (at a comparable moment in time), using only those cars to calculate travel time distributions does not lead to useful results. Instead, the route should be split up into segments, travel time distributions are computed for each segment and then combined into a travel time distribution for the entire route.

Our analysis shows that in this manner tavel time distributions can be calculated based on historical data that are close to the actual travel time ditribution on the day for which the distribution is made. Our analysis also shows that the results vary strongly, depending on: the level of granularity of the time frame for which the prediction is made; the historical data on which the prediction is based; the extent to which the travel time of a segment of a route is assumed to be related to that of the previous segment.

As timeframes, we investigated timeframes of 5 minutes, 20 minutes and 5 timeframes per day (morning rush hour, day, evening rush hour, evening and night). We concluded that short timeframes do not lead to more accurate travel time distributions.

As historical data, we used data from the same weekday of one week, four weeks and three months ago, i.e.: one, four of twelve days of historical data. We concluded that taking more historical days to make the prediction does not improve the prediction in terms of the probability that the predicted distribution and the actual distribution overlap.

As segment dependency, we investigated a low and a high segment dependence. We concluded that a high segment dependency, although it leads to less precise estimates, creates a prediction that is closer to the actual travel time distribution. 

\section*{Acknowledgement}
The work descirbed in this paper was done as part of the DAIPEX project funded by Dinalog.


\begin{thebibliography}{30}
\providecommand{\natexlab}[1]{#1}
\providecommand{\url}[1]{\texttt{#1}}
\providecommand{\urlprefix}{URL }
\expandafter\ifx\csname urlstyle\endcsname\relax
  \providecommand{\doi}[1]{doi:\discretionary{}{}{}#1}\else
  \providecommand{\doi}[1]{doi:\discretionary{}{}{}\begingroup
  \urlstyle{rm}\url{#1}\endgroup}\fi
\providecommand{\bibinfo}[2]{#2}

\bibitem[{Korteweg(????)}]{dutchpolicy}
\bibinfo{author}{J.~Korteweg}, \bibinfo{title}{Dutch Mobility Policy Document
  in a European Context} .

\bibitem[{Daganzo and Daganzo(1997)}]{daganzo1997fundamentals}
\bibinfo{author}{C.~Daganzo}, \bibinfo{author}{C.~Daganzo},
  \bibinfo{title}{Fundamentals of transportation and traffic operations},
  vol.~\bibinfo{volume}{30}, \bibinfo{publisher}{Pergamon Oxford},
  \bibinfo{year}{1997}.

\bibitem[{Hoogendoorn and Bovy(2001)}]{Hoogendoorn2001state}
\bibinfo{author}{S.~Hoogendoorn}, \bibinfo{author}{P.~Bovy},
  \bibinfo{title}{State-of-the-art of vehicular traffic flow modelling},
  \bibinfo{journal}{Proceedings of the Institution of Mechanical Engineers,
  Part I: Journal of Systems and Control Engineering}
  \bibinfo{volume}{215}~(\bibinfo{number}{4}) (\bibinfo{year}{2001})
  \bibinfo{pages}{283--303}.

\bibitem[{{van Lint} et~al.(2005){van Lint}, Hoogendoorn, and {van
  Zuylen}}]{vanLint2005advanced}
\bibinfo{author}{J.~{van Lint}}, \bibinfo{author}{S.~Hoogendoorn},
  \bibinfo{author}{H.~{van Zuylen}}, \bibinfo{title}{Accurate freeway travel
  time prediction with state-space neural networks under missing data},
  \bibinfo{journal}{Transportation Research Part C: Emerging Technologies}
  \bibinfo{volume}{13}~(\bibinfo{number}{5}) (\bibinfo{year}{2005})
  \bibinfo{pages}{347--369}.

\bibitem[{Min and Wynter(2011)}]{Min2011real}
\bibinfo{author}{W.~Min}, \bibinfo{author}{L.~Wynter},
  \bibinfo{title}{Real-time road traffic prediction with spatio-temporal
  correlations}, \bibinfo{journal}{Transportation Research Part C: Emerging
  Technologies} \bibinfo{volume}{19}~(\bibinfo{number}{4})
  (\bibinfo{year}{2011}) \bibinfo{pages}{606 -- 616}.

\bibitem[{Williams(2001)}]{Williams2001multivariate}
\bibinfo{author}{B.~Williams}, \bibinfo{title}{Multivariate Vehicular Traffic
  Flow Prediction: Evaluation of ARIMAX Modeling},
  \bibinfo{journal}{Transportation Research Record: Journal of the
  Transportation Research Board} \bibinfo{volume}{1776} (\bibinfo{year}{2001})
  \bibinfo{pages}{194--200}.

\bibitem[{Ishak and Al-Deek(2002)}]{Ishak2002performance}
\bibinfo{author}{S.~Ishak}, \bibinfo{author}{H.~Al-Deek},
  \bibinfo{title}{Performance Evaluation of Short-Term Time-Series Traffic
  Prediction Model}, \bibinfo{journal}{Journal of Transportation Engineering}
  \bibinfo{volume}{128}.

\bibitem[{Chen and Rakha(2014)}]{Chen2014real}
\bibinfo{author}{H.~Chen}, \bibinfo{author}{H.~Rakha},
  \bibinfo{title}{Real-time travel time prediction using particle filtering
  with a non-explicit state-transition model}, \bibinfo{journal}{Transportation
  Research Part C: Emerging Technologies} \bibinfo{volume}{43}
  (\bibinfo{year}{2014}) \bibinfo{pages}{112--126}.

\bibitem[{Fei et~al.(2011)Fei, Lu, and Liu}]{Fei2011bayesian}
\bibinfo{author}{X.~Fei}, \bibinfo{author}{C.~Lu}, \bibinfo{author}{K.~Liu},
  \bibinfo{title}{A bayesian dynamic linear model approach for real-time
  short-term freeway travel time prediction}, \bibinfo{journal}{Transportation
  Research Part C: Emerging Technologies}
  \bibinfo{volume}{19}~(\bibinfo{number}{6}) (\bibinfo{year}{2011})
  \bibinfo{pages}{1306--1318}.

\bibitem[{Dia and Berkum(2001)}]{Dia2001object}
\bibinfo{author}{H.~Dia}, \bibinfo{author}{E.~Berkum}, \bibinfo{title}{An
  object oriented neural network approach to short-term traffic forecasting},
  \bibinfo{journal}{European Journal of Operation Research}
  \bibinfo{volume}{131}~(\bibinfo{number}{2}) (\bibinfo{year}{2001})
  \bibinfo{pages}{253--261}.

\bibitem[{Hinsbergen et~al.(2011)Hinsbergen, Hegyi, {van Lint}, and
  Zuylen}]{Hinsbergen2011bayesian}
\bibinfo{author}{C.~Hinsbergen}, \bibinfo{author}{A.~Hegyi},
  \bibinfo{author}{J.~{van Lint}}, \bibinfo{author}{H.~Zuylen},
  \bibinfo{title}{Bayesian neural networks for the prediction of stochastic
  travel times in urban networks}, \bibinfo{journal}{IET Intel. Transport
  Syst.} \bibinfo{volume}{5} (\bibinfo{year}{2011}) \bibinfo{pages}{259--265}.

\bibitem[{Elhenawy et~al.(2014)Elhenawy, Chen, and Rakha}]{Elhenawy2014dynamic}
\bibinfo{author}{M.~Elhenawy}, \bibinfo{author}{H.~Chen},
  \bibinfo{author}{H.~Rakha}, \bibinfo{title}{Dynamic travel time prediction
  using data clustering and genetic programming},
  \bibinfo{journal}{Transportation Research Part C: Emerging Technologies}
  \bibinfo{volume}{42} (\bibinfo{year}{2014}) \bibinfo{pages}{82--98}.

\bibitem[{Nanthawichit et~al.(2003)Nanthawichit, Nakatsuji, and
  Suzuki}]{Nanthawichit2003application}
\bibinfo{author}{C.~Nanthawichit}, \bibinfo{author}{T.~Nakatsuji},
  \bibinfo{author}{H.~Suzuki}, \bibinfo{title}{Application of Probe-Vehicle
  Data for Real-Time Traffic-State Estimation and Short-Term Travel-Time
  Prediction on a Freeway}, \bibinfo{journal}{Transportation Research Record:
  Journal of the Transportation Research Board} \bibinfo{volume}{1855}
  (\bibinfo{year}{2003}) \bibinfo{pages}{49--59}.

\bibitem[{Kim and Mahmassani(2014)}]{Kim2014finite}
\bibinfo{author}{J.~Kim}, \bibinfo{author}{H.~Mahmassani}, \bibinfo{title}{A
  finite mixture model of vehicle-to-vehicle and day-to-day variability of
  traffic network travel times}, \bibinfo{journal}{Transportation Research Part
  C: Emerging Technologies} \bibinfo{volume}{46} (\bibinfo{year}{2014})
  \bibinfo{pages}{83 -- 97}.

\bibitem[{Susilawati et~al.(2014)Susilawati, Taylor, and
  Somenahalli}]{Susilawati2014urban}
\bibinfo{author}{Susilawati}, \bibinfo{author}{M.~Taylor},
  \bibinfo{author}{S.~Somenahalli}, \bibinfo{title}{Urban Arterial Road Travel
  Time Variability Modeling Using Burr Regression}, in:
  \bibinfo{booktitle}{Proceedings of the Transportation Research Board 93rd
  Annual Meeting}, \bibinfo{pages}{16}, \bibinfo{year}{2014}.

\bibitem[{Kaparias and Bell(2008)}]{Kaparias2008new}
\bibinfo{author}{I.~Kaparias}, \bibinfo{author}{H.~Bell, M~andBelzner},
  \bibinfo{title}{A New Measure of Travel Time Reliability for In-Vehicle
  Navigation Systems}, \bibinfo{journal}{Journal of Intelligent Transportation
  Systems} \bibinfo{volume}{12}~(\bibinfo{number}{4}) (\bibinfo{year}{2008})
  \bibinfo{pages}{202--211}.

\bibitem[{Eisele et~al.(2015)Eisele, Naik, and Rilett}]{Eisele2015estimating}
\bibinfo{author}{W.~Eisele}, \bibinfo{author}{B.~Naik},
  \bibinfo{author}{L.~Rilett}, \bibinfo{title}{Estimating route travel time
  reliability from simultaneously collected link and route vehicle probe data
  and roadway sensor data}, \bibinfo{journal}{International Journal of Urban
  Sciences} \bibinfo{volume}{19}~(\bibinfo{number}{3}) (\bibinfo{year}{2015})
  \bibinfo{pages}{286--304}.

\bibitem[{Yeon et~al.(2008)Yeon, Elefteriadou, and
  Lawphongpanich}]{Yeon2008breakdown}
\bibinfo{author}{J.~Yeon}, \bibinfo{author}{L.~Elefteriadou},
  \bibinfo{author}{S.~Lawphongpanich}, \bibinfo{title}{Travel time estimation
  on a freeway using Discrete Time Markov Chains},
  \bibinfo{journal}{Transportation Research Part B: Methodological}
  \bibinfo{volume}{42}~(\bibinfo{number}{4}) (\bibinfo{year}{2008})
  \bibinfo{pages}{325--338}.

\bibitem[{Ramezani and Geroliminis(2012)}]{Ramezani2012estimation}
\bibinfo{author}{M.~Ramezani}, \bibinfo{author}{N.~Geroliminis},
  \bibinfo{title}{On the estimation of arterial route travel time distribution
  with Markov chains}, \bibinfo{journal}{Transportation Research Part B:
  Methodological} \bibinfo{volume}{46} (\bibinfo{year}{2012})
  \bibinfo{pages}{1576--1590}.

\bibitem[{Rahmani et~al.(2015)Rahmani, Jenelius, and
  Koutsopoulos}]{Rahmani2015floating}
\bibinfo{author}{M.~Rahmani}, \bibinfo{author}{E.~Jenelius},
  \bibinfo{author}{H.~Koutsopoulos}, \bibinfo{title}{Non-parametric estimation
  of route travel time distributions from low-frequency floating car data},
  \bibinfo{journal}{Transportation Research Part C: Emerging Technologies}
  \bibinfo{volume}{58} (\bibinfo{year}{2015}) \bibinfo{pages}{343--362}.

\bibitem[{Hernan et~al.(2016)Hernan, Ha, and Qing}]{Caceres2016estimating}
\bibinfo{author}{C.~Hernan}, \bibinfo{author}{H.~Ha},
  \bibinfo{author}{H.~Qing}, \bibinfo{title}{Estimating freeway route travel
  time distributions with consideration to time-of-day, inclement weather, and
  traffic incidents}, \bibinfo{journal}{Journal of Advanced Transportation}
  \bibinfo{volume}{50}~(\bibinfo{number}{6}) (\bibinfo{year}{2016})
  \bibinfo{pages}{967--987}.

\bibitem[{Sun et~al.(2008)Sun, Yang, and Mahmassani}]{Sun2008travel}
\bibinfo{author}{L.~Sun}, \bibinfo{author}{J.~Yang},
  \bibinfo{author}{H.~Mahmassani}, \bibinfo{title}{Travel time estimation based
  on piecewise truncated quadratic speed trajectory},
  \bibinfo{journal}{Transportation Research Part A: Policy and Practice}
  \bibinfo{volume}{42}~(\bibinfo{number}{1}) (\bibinfo{year}{2008})
  \bibinfo{pages}{173--186}.

\bibitem[{Seo et~al.(2015)Seo, Kusakabe, and Asakura}]{Seo2015estimation}
\bibinfo{author}{T.~Seo}, \bibinfo{author}{T.~Kusakabe},
  \bibinfo{author}{Y.~Asakura}, \bibinfo{title}{Estimation of flow and density
  using probe vehicles with spacing measurement equipment},
  \bibinfo{journal}{Transportation Research Part C: Emerging Technologies}
  \bibinfo{volume}{53} (\bibinfo{year}{2015}) \bibinfo{pages}{134--150}.

\bibitem[{Qiu et~al.(2010)Qiu, Lu, Chow, and Shladover}]{Qiu2010estimation}
\bibinfo{author}{T.~Qiu}, \bibinfo{author}{X.~Lu}, \bibinfo{author}{A.~Chow},
  \bibinfo{author}{S.~Shladover}, \bibinfo{title}{Estimation of Freeway Traffic
  Density with Loop Detector and Probe Vehicle Data},
  \bibinfo{journal}{Transportation Research Record: Journal of the
  Transportation Research Board} \bibinfo{volume}{2178} (\bibinfo{year}{2010})
  \bibinfo{pages}{21--29}.

\bibitem[{Ou et~al.(2008)Ou, {van Lint}, and Hoogendoorn}]{Ou2008piecewise}
\bibinfo{author}{Q.~Ou}, \bibinfo{author}{J.~{van Lint}},
  \bibinfo{author}{S.~Hoogendoorn}, \bibinfo{title}{Piecewise Inverse Speed
  Correction by Using Individual Travel Times},
  \bibinfo{journal}{Transportation Research Record: Journal of the
  Transportation Research Board} \bibinfo{volume}{2049} (\bibinfo{year}{2008})
  \bibinfo{pages}{92--102}.

\bibitem[{Herrera et~al.(2010)Herrera, Work, Herring, Ban, Jacobson, and
  Bayen}]{herrera2010evaluation}
\bibinfo{author}{J.~Herrera}, \bibinfo{author}{D.~Work},
  \bibinfo{author}{R.~Herring}, \bibinfo{author}{X.~Ban},
  \bibinfo{author}{Q.~Jacobson}, \bibinfo{author}{A.~Bayen},
  \bibinfo{title}{Evaluation of traffic data obtained via GPS-enabled mobile
  phones: The Mobile Century field experiment},
  \bibinfo{journal}{Transportation Research Part C: Emerging Technologies}
  \bibinfo{volume}{18}~(\bibinfo{number}{4}) (\bibinfo{year}{2010})
  \bibinfo{pages}{568--583}.

\bibitem[{Treiber and Kesting(2013)}]{treiber2013traffic}
\bibinfo{author}{M.~Treiber}, \bibinfo{author}{A.~Kesting},
  \bibinfo{title}{Traffic flow dynamics}, \bibinfo{journal}{Traffic Flow
  Dynamics: Data, Models and Simulation, Springer-Verlag Berlin Heidelberg} .

\bibitem[{Kerner(2012)}]{kerner2012physics}
\bibinfo{author}{B.~Kerner}, \bibinfo{title}{The physics of traffic: empirical
  freeway pattern features, engineering applications, and theory},
  \bibinfo{publisher}{Springer}, \bibinfo{year}{2012}.

\bibitem[{Treiber(2015)}]{extralane}
\bibinfo{author}{M.~Treiber}, \bibinfo{title}{Traffic Flow Dynamics{,} Traffic
  States}, \urlprefix\url{http://www.h2063376.stratoserver.net/trafficstates/},
  \bibinfo{note}{accessed: 30 July 2018}, \bibinfo{year}{2015}.

\bibitem[{{van Lint} et~al.(2008){van Lint}, {van Zuylen}, and
  Tu}]{van2008travel}
\bibinfo{author}{J.~{van Lint}}, \bibinfo{author}{H.~{van Zuylen}},
  \bibinfo{author}{H.~Tu}, \bibinfo{title}{Travel time unreliability on
  freeways: Why measures based on variance tell only half the story},
  \bibinfo{journal}{Transportation Research Part A: Policy and Practice}
  \bibinfo{volume}{42}~(\bibinfo{number}{1}) (\bibinfo{year}{2008})
  \bibinfo{pages}{258--277}.

\end{thebibliography}
\end{document}